\documentclass[12pt,english,aps,pra]{revtex4}
\usepackage[T1]{fontenc}
\usepackage[latin9]{inputenc}
\usepackage{color}
\usepackage{units}
\usepackage{amsmath}
\usepackage{amssymb}
\usepackage{graphicx}
\usepackage{esint}

\makeatletter
\@ifundefined{textcolor}{}
{%
 \definecolor{BLACK}{gray}{0}
 \definecolor{WHITE}{gray}{1}
 \definecolor{RED}{rgb}{1,0,0}
 \definecolor{GREEN}{rgb}{0,1,0}
 \definecolor{BLUE}{rgb}{0,0,1}
 \definecolor{CYAN}{cmyk}{1,0,0,0}
 \definecolor{MAGENTA}{cmyk}{0,1,0,0}
 \definecolor{YELLOW}{cmyk}{0,0,1,0}
 }

\usepackage{mathrsfs}
\usepackage{bbm}

\makeatother

\usepackage{babel}

\makeatother

\usepackage{babel}

\newcommand{\eq}[2]{\begin{equation} #2 \label{#1} \end{equation}}
\newcommand{\paren}[1]{\left(#1\right)}

\begin{document}

\title{Multipartite  Quantum Entanglement Evolution in Photosynthetic Complexes}

\author{Jing Zhu and Sabre Kais}

\address{Department of Chemistry and Birck Nanotechnology Center, Purdue University,
West Lafayette, IN 47907, USA}

\author{Al\'{a}n Aspuru-Guzik}

\address{Department of Chemistry and Chemical Biology, Harvard University,
12 Oxford Street, Cambridge, MA 02138, USA}

\author{Sam Rodriques, Ben Brock and Peter J. Love}

\thanks{Corresponding author, plove@haverford.edu}

\address{Department of Physics, Haverford College, Haverford, PA 19041}

\begin{abstract}
We investigate the evolution of entanglement in the Fenna-Matthew-Olson (FMO) complex based on simulations using the scaled hierarchical equations of motion (HEOM) approach.  We examine the role of entanglement in the FMO complex by direct computation of the convex roof.  We use monogamy to give a lower bound for entanglement and obtain an upper bound from the evaluation of the convex roof.  Examination of bipartite measures for all possible bipartitions provides a complete picture of the multipartite entanglement. Our results support the hypothesis that entanglement is maximum primary along the two distinct electronic energy transfer pathways. In addition, we note that the structure of multipartite entanglement is quite simple, suggesting that there are constraints on the mixed state entanglement beyond those due to monogamy.

\end{abstract}
\maketitle

\section{Introduction}

Photosynthesis is one of the most common phenomena in nature. However, the details of photosynthetic processes are still under
investigation. Recent experimental results show that long lived quantum coherences are present in various photosynthetic complexes~\cite{Andrew-book,Scholes-2010,NatureReview}. One such protein complex, the Fenna-Matthews-Olson (FMO) complex from green sulphur bacteria \cite{FENNA1975}, has attracted a great deal of experimental and theoretical attention due to its intermediate role in energy transport. The FMO complex acts as a molecular wire, transferring the excitation energy from the light-harvesting complex (LHC) to the reaction center (RC) \cite{FENNA1975,Li1997,Camara-Artigas2003,Cheng-Fleming-2009}. In 2007, Engel et al \cite{Engel-2007} observed long-lasting quantum beating over a time scale of hundreds of femtoseconds by two-dimensional nonlinear spectroscopy. Evidence for quantum beating, and therefore long lived quantum coherence, was also found at room temperature \cite{Engel2}.

The transport of electronic excitations through the protein complex of FMO is an example of energy transport in an open quantum system. The oscillations of the nuclear positions provide a bath or an environment for the electronic excitations. Since $2007$, several theoretical frameworks have been developed to model this phenomenon. For example, Aspuru-Guzik et al \cite{Alan-2008,Alan-2009,Alan-JPC-2009} introduced a non-Markov approximation based on the Lindblad formalism to investigate the effects on the efficiency of photosynthesis of the combination of quantum coherence and environmental interaction. Meanwhile, Ishizaki and coworkers \cite{Fleming-JCP-2009,Fleming-PNAS-2009} utilized the hierarchical equations of motion (HEOM) approach to reproduce  successfully the population beating in the FMO complex at both cryogenic and physiological temperature. More recently, Zhu and coworkers introduced the scaled HEOM approach for studying the robustness and quantum coherence in the FMO complex~ \cite{Ourpaper,Shi2009a}. The scaled HEOM approach has been shown to provide reliable simulation results with considerable reduction in computational requirements. Using the HEOM equations, Rebentrost and Aspuru-Guzik showed  that the non-Markovianity of the system is near-maximal for physiological conditions \cite{Alan2011Com}. Recently, many other approaches for the numerical computation of the time evolution and quantum features of this system have made FMO a target for benchmarking of methods for simulating open quantum systems \cite{Reichman2011,PlenioDMRG2,coker2010,coker2011,Mazziotti2011,Mazziotti2011a,Thorwart2011,Lloyd2011,Lloyd2011a,Lloyd2011b,cao2010,cao2011,Silbey2010,Silbey2011,Ratner2011,Mukamel2010a}.

Besides the modeling of population and coherence observed in experiment, these models also enable computation of the time evolution of entanglement~\cite{Horodecki:2009p8660,Kais2007}. The first study of entanglement in biological excitons was~\cite{Thorwart:2009p8612}, which studied the dynamics of the negativity~\cite{Peres:1996p8927,Horodecki:1996p8928} for a pair of chromophores coupled to a non-Markovian environment. Subsequent studies considered more chromophores, different excitation mechanisms and different entanglement measures. We briefly review this work here, for a more complete overview we refer the reader to a recent review~\cite{Whaley:2010p8611}. In a recent study, Mukamel made a distinction between some apparent entanglement effects associated with the linear response, which can be eliminated by a coordinate transformation, and genuine entanglement that is fundamentally quantum in nature \cite{Mukamel2010}. Recently, Engel et al found a direct evidence of quantum transport in the FMO complex \cite{engel2011}.

In~\cite{Sarovar:2010p6945} two measures of entanglement relevant to FMO are defined. The first measure is the concurrence between chromophore $i$ and chromophore $j$. The concurrence is a well-known measure of entanglement between two two-level systems, and can be computed in closed form even for mixed states, and in the case of a density matrix restricted to the single exciton subspace takes the simple form $C_{ij} = 2|\rho_{ij}|$~\cite{Sarovar:2010p6945,Wootters:1998p6991}.  The second measure defined was a global measure related to the relative entropy of entanglement, defined by;
\begin{equation}\label{seREE}
E[\rho]=-\sum_{i=1}^N\ln\rho_{ii} -S(\rho)	
\end{equation}
where $S(\rho ) = -{\rm Tr}\rho \ln \rho$ is the von Neumann entropy of the state $\rho$.  This measure is the relative entropy of entanglement specialized to the case where states only have support in the zero and one exciton subspace.  The definition of the relative entropy of entanglement is
\begin{equation}
E[\rho] = \min_\sigma {\rm Tr}(\rho\ln\rho - \rho\ln\sigma)
\end{equation}
where the minimization is taken over all separable states $\sigma$. In the case of states restricted to zero or one excitons, the set of separable states becomes simply the set of diagonal density matrices, and so this minimization can be performed exactly, yielding the expression~(\ref{seREE}). We refer the reader to the supplementary materials of~\cite{Sarovar:2010p6945} for more details. Both of the measures computed in~\cite{Sarovar:2010p6945} rely on the fact that, in the single exciton subspace, coherence (meaning nonzero off diagonal elements of the density matrix in the standard basis) is necessary and sufficient for entanglement. Both concurrence, the relative entropy of entanglement and an entanglement witness introduced in~\cite{Sarovar:2010p6945} show this clearly.

We introduce the notation that the bipartition of a system into subsystems $A$ and $B$ is denoted $A|B$, and when a subsystem consists of a set of chromophores we indicate it by a string of labels, so $12|367$ is the bipartition of the subsystem composed of chromophores one and two $(12)$, and the subsystem composed of chromophores three, six and seven $(367)$.

The two measures considered in~\cite{Sarovar:2010p6945} were computed for an initial excitation at site one or six, at both $77K$ and $300K$, to probe both physiological conditions and the conditions of ultrafast spectroscopy experiments. For the system initialized with an exciton at site 1, they show the pairwise entanglement  $1|2$, $1|3$, $1|5$ and also the pairwise entanglement $3|4$. Finite entanglement was found between all pairs of chromophores in~\cite{Sarovar:2010p6945} - over distances comparable to the size of the FMO complex - $\leq 30$\AA.

The logarithmic negativity is the only measure that is readily computable for all states, and in the case of states restricted to the single exciton subspace it may be computed across any cut of the set of seven chromophores into two subsets~\cite{logneg,Caruso:2010p8915,Caruso:2009p8916}. Caruso {\em et al.} computed the logarithmic negativity across six cuts $1|234567$, $12|34567$, $123|4567$, $1234|567$, $12345|67$ and $123456|7$ in a simulation in which a single excitation was injected into site one~\cite{Caruso:2010p8915}. The entanglement of site one with the rest $1|234567$ exhibited the largest peak value, with large oscillations taking it below the entanglements across the other cuts. This may be understood as the generation of entanglement from the delocalization of the injected exciton across the complex. In subsequent work, the logarithmic negativity was also computed (across the same cuts) for simulations in which direct injection of a single exciton is replaced by simulation of thermal injection and laser excitation. In the case of thermal injection the entanglement is reduced by a factor of roughly $50$, concomitant with a suppression of coherent oscillations. In the case of simulated laser excitation a large pulse of entanglement is observed, lasting about $0.15~{\rm ps}$.

In~\cite{Fassioli:2010p8617} Fassioli et al move from consideration of the presence of entanglement in models of FMO to characterization of its functional role in transport. It is in this context that the variety of entanglement studies carried out could connect with functionality and delocalization ideas from physical chemistry. Those authors introduce an entanglement yield, based on an entanglement measure which is a sum of the squared concurrences or ``tangles'' (defined below) over all pairs of chromophores.
\begin{equation}
E_T = \sum_{m,n>m} \tau(\rho_{m,n})
\end{equation}
Because of monogamy of entanglement their measure is bounded above by a sum of the tangles of each chromophore with the rest.
\begin{equation}
E_T  \leq \frac{1}{2}\sum_{n} \tau(\rho_n)
\end{equation}
This upper bound is equal to $7/2$ times the Meyer-Wallach measure for the seven chromophore system~\cite{Meyer:2001p8917}.  Interestingly, those authors point out a connection of this measure, and hence of the Meyer-Wallach measure, to a measure commonly used by the physical chemistry community of exciton delocalization: the inverse participation ratio~\cite{Meier:1997p8934}.

To make a connection between entanglement and transport Fassioli {\em et al.}~\cite{Fassioli:2010p8617} define an entanglement yield - the integral of the entanglement (as given by a sum of pairwise tangles) weighted by the probability density for exciton absorption by the reaction center. This quantity is normalized by the quantum yield: the total probability that the exciton is trapped by the reaction center.  The  contributions to this quantity were divided into donor-donor, donor-acceptor contributions, where chromophores 1, 2 and 5, 6 are  designated donors and chromophores 3 and 4 are acceptors. This study showed that entanglement peaks on a timescale relevant for transport, for simulations in which the initial exciton is localized on site one or site six. In particular those authors observe an inverse relationship between entanglement among donor sites and quantum efficiency, suggesting that entanglement among the donor chromophores (1,2 and 5,6) may be tuned to achieve the desired quantum efficiency. The authors of~\cite{Fassioli:2010p8617} also introduce the idea of direct and indirect pathways - an indirect pathway involving transfer through chromophore seven. The connection between entanglement and transport was also made clear by the work of~\cite{addone} in which it was shown that a high probability of exciton transfer was only achieved for large values of the entanglement.

In~\cite{Bradler:2010p7069} a number of distinct measures of quantum correlation were computed: the quantum mutual information, quantum discord and single-excitation relative entropy of entanglement with respect to bipartite cuts $3|16$, $12|3$ and $3|124567$. These authors extended the work of ~\cite{Sarovar:2010p6945} by proving a simple formula for the relative entropy of entanglement across any bipartite cut for states restricted to the single exciton subspace.

It is the goal of the present work to provide a more complete picture of entanglement evolution during exciton transport. We also wish to further investigate the relationship of entanglement to the different transport pathways in the context of the HEOM model presented below. The paper is organized as the follows. In Section~\ref{HEOM} the detailed theoretical framework of the scaled HEOM approach is introduced.  In Section~\ref{ent} the method used to compute the convex roof and hence obtain the entanglement is given. Section~\ref{res} contains our entanglement calculations. We use the monogamy bounds in order to validate our convex roof method - the monogamy bounds provide a lower bound on entanglement and our convex roof calculations provide an upper bound. We compute bipartite measures of entanglement (described in detail below) for many subsystems and bipartitions of the FMO complex, including calculations for all $63$ bipartitions of the full seven chromophore system in order to provide a full picture of the multipartite entanglement present during transport. We close the paper with some conclusions and directions for future work.

\section{Method: Scaled Hierarchical Equations of Motion (HEOM)}\label{HEOM}

The structure of the FMO complex was originally analyzed by Fenna and Matthews \cite{FENNA1975}. The FMO complex consists of three identical monomers arranged in a C3 symmetric structure. Each monomer works independently in the FMO complex.  Each monomer is formed from seven bacteriochlorophylla (BChla) molecules. These molecules are the ``sites'' or ``chromophores'' referred to in the rest of the paper.  Experimental results show that site $1$ and $6$ are close to the light Harvesting complex (LHC) and site $3$ and $4$ are next to the reaction center (RC)~\cite{FENNA1975,Li1997,Camara-Artigas2003,Cheng-Fleming-2009}.

For all models used in the present paper, the Hamiltonian of the FMO complex and its interaction with the environment is taken to be:
\begin{align}
\mathcal{H} & =\mathcal{H_{S}}+\mathcal{H}_{B}+\mathcal{H}_{SB}\label{eq:htot}\\
\mathcal{H}_{S} & =\sum_{j=1}^{N}\varepsilon_{j}\,|j\rangle\langle j|+\sum_{j\neq k}J_{jk}\,\left(|j\rangle\langle k|+|k\rangle\langle j|\right)\label{eq:hs}\\
\mathcal{H}_{B} & =\sum_{j=1}^{N}\mathcal{H}_{B}^{j}=\sum_{j=1}^{N}\sum_{\xi=1}^{N_{jB}}\frac{P_{j\xi}^{2}}{2m_{j\xi}}+\frac{1}{2}m_{j\xi}\omega_{j\xi}^{2}x_{j\xi}^{2}\label{eq:hb}\\
\mathcal{H}_{SB} & =\sum_{j=1}^{N}\mathcal{H}_{SB}^{j}=-\sum_{j=1}^{N}|j\rangle\langle j|\cdot\sum_{\xi}c_{j\xi}\cdot x_{j\xi}=-\sum_{j=1}^{N}\mathcal{V}_{j}\cdot F_{j}\label{eq:hsb}\\
 & \mbox{with }\mathcal{V}_{j}=|j\rangle\langle j|\mbox{ and }F_{j}=\sum_{\xi}c_{j\xi}\cdot x_{j\xi}\nonumber
\end{align}

The terms $\mathcal{H}_{S}$, $\mathcal{H}_{B}$ and $\mathcal{H}_{SB}$ describe the Hamiltonian of the system, the bath, and the system-bath coupling respectively. The Hamiltonian is written in the single excitation subspace, so that the basis states $|j\rangle$ in Eq.~\ref{eq:hs} denotes that the $j$-th site is in its excited state and all other sites are in their ground states. The energy of site $j$ is denoted by $\varepsilon_{j}$ and $J_{jk}$ is the electronic coupling between site $j$ and $k$. $N$ is the number of sites, so that $N=7$ for the FMO complex. For the thermal bath $\mathcal{H}_{B}$, the harmonic oscillator model is applied. We assume that each site is coupled to the bath independently. The parameters $m_{j\xi}$, $\omega_{j\xi}$, $P_{j\xi}$ and $x_{j\xi}$ are mass, frequency, momentum and position operator of the harmonic bath associated with the $j$-th site respectively. The parameter $c_{j\xi}$ in Eq. \ref{eq:hsb} represents the system-bath coupling constant between the $j$-th site and $\xi$-th phonon mode. The system and bath are assumed to be decoupled at $t=0$.

We can obtain the time evolution of the system density matrix $\rho\left(t\right)$ by tracing out the bath degrees of freedom $\rho\left(t\right)={\rm Tr}_{B}\left[\rho_{tot}\left(t\right)\right]={\rm Tr}_{B}\left[e^{\nicefrac{-i\mathcal{H}t}{\hbar}}\,\rho_{tot}\left(0\right)\,e^{\nicefrac{i\mathcal{H}t}{\hbar}}\right]$.
The correlation function for a phonon bath can be written as
\begin{align}
C_{j}\left(t\right) & =\frac{1}{\pi}\intop_{-\infty}^{\infty}d\omega\cdot J_{j}\left(\omega\right)\cdot\frac{e^{-i\omega t}}{1-e^{-\beta\hbar\omega}}\label{eq:correlation}\\
J_{j}\left(\omega\right) & =\sum_{\xi}\frac{c_{j\xi}^{2}\cdot\hbar}{2m_{j\xi}\cdot\omega_{j\xi}}\delta\left(\omega-\omega_{j\xi}\right)\label{eq:spec density}
\end{align}
with $\beta=\nicefrac{1}{k_{B}T}$ . We assume that $J_{j}\left(\omega\right)$ is the same all sites, $J_{j}\left(\omega\right)=J\left(\omega\right)\;\forall\; j\mbox{s}$.  We consider the time evolution of the system density matrix both with and without environmental interaction. For the isolated system, we set $J\left(\omega\right)=0$ and the time evolution of the density matrix for the system is given by:
\begin{equation}
\frac{d}{dt}\rho\left(t\right)=-\frac{i}{\hbar}\left[\mathcal{H}_{S},\;\rho\left(t\right)\right]\label{eq:isolated}
\end{equation}
One approach to the computation of the time evolution of the system density matrix is the hierarchical equation of motion (HEOM) approach, originally developed by Ishizaki and Fleming \cite{Fleming-PNAS-2009}. We use the scaled HEOM approach for reasons of computational efficiency~\cite{Shi2009a,Ourpaper}.

In the scaled HEOM approach, the original spectral density function $J\left(\omega\right)$ (Eq. \ref{eq:spec density}) is replaced by a Drude spectral density function $J\left(\omega\right)=\frac{2\lambda\gamma}{\hbar}\frac{\omega}{\omega^{2}+\gamma^{2}}$ where $\lambda$ is the reorganization energy and $\gamma$ is the Drude decay constant. Then the correlation function in Eq. \ref{eq:correlation} can be expanded as
\begin{align*}
C_{j}\left(t>0\right) & =\sum_{k=0}^{\infty}c_{k}\cdot e^{-v_{k}t}
\end{align*}
with $v_{o}=\gamma$, which is the Drude decay constant, $v_{k}=\frac{2k\pi}{\beta\hbar}$ when $k\geqslant1$ and $v_{k}$ is known as the Matsuraba frequency.
The constants $c_{k}$ are given by
\begin{eqnarray*}
c_{0} & = & \frac{\eta\gamma}{2}\left[\cot\left(\frac{\beta\hbar\gamma}{2}\right)-i\right]\\
c_{k} & = & \frac{2\eta\gamma}{\beta\hbar}\cdot\frac{v_{k}}{v_{k}^{2}-\gamma^{2}}\;\; for\,\, k\geqslant1
\end{eqnarray*}

Using the scaled approach developed by Shi and coworkers \cite{Shi2009a} and applying the Ishizaki-Tanimura truncating scheme \cite{Tanimura,Tanimura2} to the density matrix, the scaled density operator becomes:
\begin{multline}
\frac{d}{dt}\rho_{\boldsymbol{n}}=-\frac{i}{\hbar}\left[\mathcal{H}_{S},\;\rho_{\boldsymbol{n}}\right]-\sum_{j=1}^{N}\sum_{k=0}^{K}n_{jk}v_{k}\cdot\rho_{\boldsymbol{n}}-i\sum_{j=1}^{N}\sqrt{\left(n_{jk}+1\right)\left|c_{k}\right|}\,\left[\mathcal{V}_{j},\;\sum_{k}\rho_{\boldsymbol{n_{jk}^{+}}}\right]\\
-\sum_{j=1}^{N}\sum_{m=K+1}^{\infty}\frac{c_{jm}}{v_{jm}}\cdot\left[\mathcal{V}_{j},\,\left[\mathcal{V}_{j},\,\rho_{\boldsymbol{n}}\right]\right]-i\sum_{j=1}^{N}\sum_{k=0}^{K}\sqrt{\nicefrac{n_{jk}}{\left|c_{k}\right|}}\;\left(c_{k}\mathcal{V}_{j}\,\rho_{\boldsymbol{n_{jk}^{-}}}-c_{k}^{*}\rho_{\boldsymbol{n_{jk}^{-}}}\mathcal{V}_{j}\right)\label{eq:final HEOM}
\end{multline}
where the global index $\boldsymbol{n}$ denotes a set of nonnegative integers $\boldsymbol{n}\equiv\{n_{1},n_{2}\cdots\cdots,n_{N}\}=\{\{n_{10},n_{11}\cdots,n_{1K}\}\cdots\{n_{N0,}n_{N1}\cdots,n_{NK}\}\}$. The symbol $\boldsymbol{n_{jk}^{\pm}}$ refers to a set in which the number $n_{jk}$ is modified to $n_{jk}\pm1$ in the global index $\boldsymbol{n}$. The sum of $n_{jk}$ is called the tier ($\mathcal{N}$), $\mathcal{N}=\sum_{j,k}n_{jk}$. The global index $\boldsymbol{n}$ labels a set of density matrices in which $\rho_{\boldsymbol{0}}=\rho_{\{\{0,0,\cdots,0\}\cdots\cdots\{0,0,\cdots,0\}\}}$ is the system reduced density operator (RDO), and all others are considered as auxiliary density operators (ADOs). Although the RDO is the most important operator, the ADOs contain corrections to the system-bath interaction, arising from the non-equilibrium treatment of the bath. $K$ is the truncation level for the correlation function (Matsuraba frequency and constant $c_{k}$) and the cutoff for the tier of ADOs was set at $\mathcal{N}_{c}$. The scaled approach guarantees that all elements in the ADOs decay to zero for the upper levels in the hierarchy, while the Ishizaki-Tanimura truncating scheme decreases the truncation error. For a detailed derivation of this approach we refer the reader to~\cite{Ourpaper}. We make use of the same parameters as~\cite{Ourpaper}, and we set the truncation levels $K=0$ and cutoff tier of ADOs $\mathcal{N}_{c}=4$. The reorganization energy and Drude decay constant are $\lambda_{j}=\lambda=35\;\unit{cm^{-1}}$ and $\gamma_{j}^{-1}=\gamma^{-1}=50\;\unit{fs}$.

By numerically integrating the differential equation Eq. \ref{eq:final HEOM}  using Mathematica, we calculated the density matrix of each time step during the evolution for $2500\unit{fs}$ with a time step of $2~{\rm fs}$. We performed simulations with two different initial states: site $1$ initially exited and site $6$ initially excited. The time series of the system density matrix so obtained is the data from which we calculate the entanglement between various different parts of the FMO complex. Before describing the results of those calculations, we first describe the method by which we compute entanglement measures for the mixed states of the seven chromophore system.

\section{Entanglement analysis}\label{ent}

The FMO complex, considered as an assembly of seven chromophores, is a multipartite quantum system. As such, useful information about quantum correlations is obtained by computing the bipartite entanglement across any of the cuts that divide the seven chromophores into two subsystems. Similarly if we take the state of any subsystem of the FMO complex we can compute the entanglement across any cut of the reduced state of that subsystem.

The measures we compute in the present paper are bipartite - they determine a measure of the entanglement between two subsystems of the 7-chromophore system. Each measure alone only contains information concerning the bipartite entanglement across the bipartition. However, the nature of multipartite entanglement in the system is given by the bipartite entanglement across all possible bipartitions (see, for example, ~\cite{Horodecki:2009p8660}, p. 890). One may therefore construct multipartite measures from multiple bipartite measures. Meyer and Wallach's ``Global'' measure of entanglement is defined as a sum of bipartite measures (an average entanglement of each subsystem with the rest). Scott~\cite{Scott:2004p8922} and Love~\cite{Love:2007p8918} both generalized Meyer and Wallach's measure to include information from further bipartitions in various averages. The first case in which interesting multiparite entanglement may occur is the case of three two-level systems. In this case a multipartite measure, the tangle, may be defined~\cite{Coffman:2000p8919}. This first example of a multipartite measure may again be expressed as a difference of bipartite measures computed for different subsystems and bipartitions.

There are $63$ distinct bipartitions of the $7$ chromophores of FMO. The bipartite entanglement across all these measures contains all multipartite entanglement information about the full system. Ideally, one would compute all of these measures to obtain a complete picture of the correlations present among subsystems.  Instead one may take subsystems and compute the entanglement across bipartitions of the subsystems. For example, by computing the entanglement between all pairs of chromophores. However, as Table~\ref{comb} shows, this leads to a large number of subsystems, and a large number of bipartitions for each subsystem.

\begin{table}
  \centering
  \begin{tabular}{|l|l|l|l|ll}
\hline
$m$ &  $7\choose m$ & Cuts & Total Measures\\
\hline
   2 & 21 &1& 21\\
\hline
    3 & 35 &3& 105\\
\hline
    4 & 35 &7& $245$\\
\hline
    5 & 21 &15&$315$\\
\hline
    6 & 7 &31&$217$\\
\hline
    7 & 1 &63& $63$ \\
\hline
\end{tabular}
  \caption{Subsystems and bipartite cuts relevant to the FMO system. One may take a subsystem reduced density matrix of any $m\leq7$ and consider all the bipartite cuts of each subsystem. This leads to a combinatoric explosion of different bipartite measures. Evidently it would be simpler to consider all cuts of the total system. We perform such convex roof calculations for the full seven chromophore system by restricting the convex roof to the single exciton manifold.}\label{comb}
\end{table}

Evidently, averaging together information from multiple bipartitions implies a loss of information, and in the present paper we simply display the measures corresponding to each bipartition directly. These calculations of bipartite measures across multiple bipartitions give us information concerning the multipartite entanglement present in the FMO system.

\subsection{Entanglement measures}

The set of monotones defined in~\cite{Love:2007p8918} for pure states of $n$ qubits is:
\begin{equation}\label{measure}
\eta_{S} = \frac{2^{|S|}}{2^{|S|}-1}\left(1-{\rm Tr}(\rho_{S}^{2})\right)
\end{equation}
where $S$ is a set of $k$ two state quantum systems (usually qubits, but in the context of the present paper these are chromophores), so that $|S|=k$, and $\rho_{S}$ is the reduced density matrix of those $k$ qubits. For two qubits with $S=1$ this measure reduces to the square of the concurrence. In order to allow easy comparison with prior work computing the concurrence for these systems we compute the square root of these measures $\sqrt{\eta_S}$ for many bipartitions of various subsystems of the seven chromophore system. We also compute these measures for all bipartitions of the full seven chromophore system.

\subsection{Monogamy of entanglement}

A fascinating property distinguishing entanglement from classical correlations is monogamy. Just as the simplest example of entanglement occurs for two qubits, the simplest example of monogamy occurs for three qubits. If, among three qubits $ABC$, the qubits $A$ and $B$ are maximally entangled, then qubit $C$ cannot be entangled at all with qubits $A$ and $B$. It is instructive to consider this from the point of view of the entanglement measures~(Eq. \ref{measure}). These measures are based on subsystem purity - if qubits $ABC$ are in a pure state and $A$ and $B$ are maximally entangled then the reduced state of qubits $AB$ is pure, hence so is the reduced state of qubit $C$, and hence qubit $C$ is unentangled with qubits $A$ and $B$. In fact, this property extends  for three qubits to the case where the entanglement is not maximal. The monogamy constraint for pure states is expressed in terms of the tangles  measuring the entanglement of qubit $A$ with a subsystem $B$:
\begin{equation}\label{tangle}
\tau_{A|B} = 2(1-{\rm Tr}\rho_A^2) =  \eta_A.
\end{equation}
where $\rho_A$ is the reduce density matrix of subsystem $A$. In terms of the measures~(Eq. \ref{tangle}) we obtain:
\begin{equation}\label{mono}
\tau_{A|B} + \tau_{A|C} \leq \tau_{A|BC}
\end{equation}
This property of three qubit states was shown in~\cite{Coffman:2000p8919}, and the result for $n$ qubits was proved in~\cite{Osborne:2006p8924}:
\begin{equation}
\sum_{i=1, i\neq m}^n \tau_{m|i} \leq \tau(m|1,\dots m-1,m+1\dots n).
\end{equation}
These imply corresponding relations among the measures $\eta_S$ that are equal to tangles of one qubit $S$ with the others.

In the context of models of exciton transport that are restricted to the single exciton subspace it is worth recalling that, in the case of pure states of three qubits, it is exactly states that are superpositions of Hamming weight one basis states that saturate the monogamy bound~\cite{Coffman:2000p8919}. In fact it has been shown that pure generalized $W$ states and mixtures of pure generalized $W$ states with $|0\rangle\langle0|$ (which corresponds to states that are incoherent combinations of the  single exciton subspace and the vacuum in the models we consider here) saturate the monogamy bounds~\cite{SanKim:2008p8940}. For pure states we may therefore obtain the entanglement of each chromophore with the rest using the sum of the pairwise entanglements. However, these bounds are not known to be saturated for the mixed states of interest here. It should be noted that the entanglement properties of W-class states also enable experimental detection of entanglement in these states~\cite{addtwo}.

It is natural to ask whether monogamy holds beyond restrictions on the entanglement of single qubits to relationships between the entanglement of higher dimensional systems. Unfortunately, in general this is not the case~\cite{Ou:2006p8925}, as it can already be shown that states of qubits violate the analogous relation to~(Eq. \ref{mono}). For the single exciton manifold of W-class states a number of relations beyond monogamy are known~\cite{SanKim:2008p8940}.  The approach we take here is to determine relationships among the measures~\ref{measure}, if any, by the direct computation of the measures. It is to the technical details of the calculation of these measures for mixed states that we turn in the next subsection.

\subsection{Convex Roof Extension of Entanglement Monotones}

The measures~\ref{measure} are defined for pure states. A general mixed state of a quantum system may also be entangled, and the measures~\ref{measure} can be extended to mixed states as follows. Given a density matrix $\rho$ and its set of ensemble representations
\begin{equation}
\aleph \equiv \left\{p_{i},|\psi_{i}\rangle : \sum\limits_{i}{p_{i}|\psi_{i}\rangle\langle\psi_{i}|} = \rho \right\},
\end{equation}
any entanglement monotone $\eta\left(|\psi\rangle\right)$ on pure states can be generalized to a monotone on mixed states, $E(\rho)$, defined by
\begin{equation}
E(\rho) \equiv {\rm inf}_{\aleph} \left[\sum\limits_{i}{p_{i}\eta\left(|\psi_{i}\rangle\right)}\right]
\end{equation}
which is also an entanglement monotone. Given a density matrix $\rho = \sum\limits_{i} {p_{i}|\psi_{i}\rangle\langle\psi_{i}|}$, define
\begin{equation}
|\phi_{i}\rangle\sqrt{q_{i}}\equiv \sum\limits_{j} {U_{ij}|\psi_{j}\rangle\sqrt{p_{j}}},
\end{equation}
where the $U_{ij}$'s are elements of a unitary matrix.  It can then be shown that $\rho = \sum\limits_{i} {q_{i}|\phi_{i}\rangle\langle\phi_{i}|}$.

Since density matrices are hermitian they are always diagonalizable.  We can therefore write $\rho = V\Lambda V^{\dagger}$; this matrix product can equivalently be written as the summation $\rho = \sum\limits_{i}{\lambda_{i}|v_{i}\rangle\langle v_{i}|}$, where the $\lambda_{i}$'s are the eigenvalues of $\Lambda$ and the $|v_{i}\rangle$'s are the basis-independent orthonormal kets corresponding to the columns of $V$.  This is called the spectral ensemble of $\rho$.  It is also useful to define $\tilde{\Phi}\equiv V\Lambda^{1/2}$, so that $\tilde{\Phi}\tilde{\Phi}^{\dagger}=\rho$.  This object $\tilde{\Phi}$ contains all the information contained in a particular ensemble, and similar objects $\tilde{\Psi}\tilde{\Psi}^{\dagger}=\rho$ also correspond to ensembles.  In fact, the unitary transform given in terms of a summation above corresponds to the matrix transformation $\tilde{\Phi}U$, where $U$ is unitary.  If we define $\tilde{\Psi} = \tilde{\Phi}U$ for some unitary matrix $U$, then $\tilde{\Psi}\tilde{\Psi}^{\dagger}=\rho$.  It can further be shown that the space of ensemble representations of $\rho$ is isomorphic to the unitary group~\cite{KirkpatrickHJW}. Hence optimization over the space of ensembles can be reduced to an optimization problem over the unitary group. We give details of the parameterization of the unitary group used in our calculations in the Appendix.

\section{Results}\label{res}

In this section, we compute a number of entanglement measures for two, three, four, five and seven qubit subsystems. Our approach follows both that of~\cite{Sarovar:2010p6945}, in which pairwise entanglements were computed, and that of~\cite{Caruso:2010p8915} in which the logarithmic negativity for several partitions of the full seven chromophore system were computed. We compute the measures $\sqrt{\eta_S}$ where $\eta_S$ is defined in eqn.~\ref{measure} for bipartitions of subsystems of two, three, four and five chromophores. For these calculations the convex roof optimization was performed in the full space of density matrices of dimension $2^7$. We then compute the measures $\sqrt{\eta_S}$ for all $63$ bipartitions of the full seven chromophore system for one initial condition, restricting the convex roof optimization to the single exciton manifold for reasons of computational tractability.

\subsection{Two site subsystems}

The pairwise concurrences are a natural starting point because they can be computed exactly, and have been the subject of extensive prior study~\cite{Sarovar:2010p6945}. We compute the reduced density matrix of each of the $21$ pairs of sites and calculate the concurrence in these two-site subsystems~\cite{Wootters:1998p6991}. For the case in which site $1$ was initially excited the coherent oscillations of population occur mainly between sites $1$ and $2$ before the energy is transferred to sites $3$ and $4$~\cite{Fleming-PNAS-2009,Ourpaper}. As a result of these coherent oscillations there is large pairwise entanglement between site $1|2$ \cite{Sarovar:2010p6945}. In the work of~\cite{Sarovar:2010p6945}, for times <900 (500) fs at 77K (300K) these measures are ordered: $1|2>1|3>1|5>3|4$. For the system of~\cite{Sarovar:2010p6945} initialized with a single exciton at site 6 the entanglements 4|5, 4|7, 5|6, 3|4 are computed. For times < 100 fs these are ordered $5|6>4|5>4|7>3|4$.

\begin{figure}
\begin{center}
\includegraphics[width=0.75\textwidth]{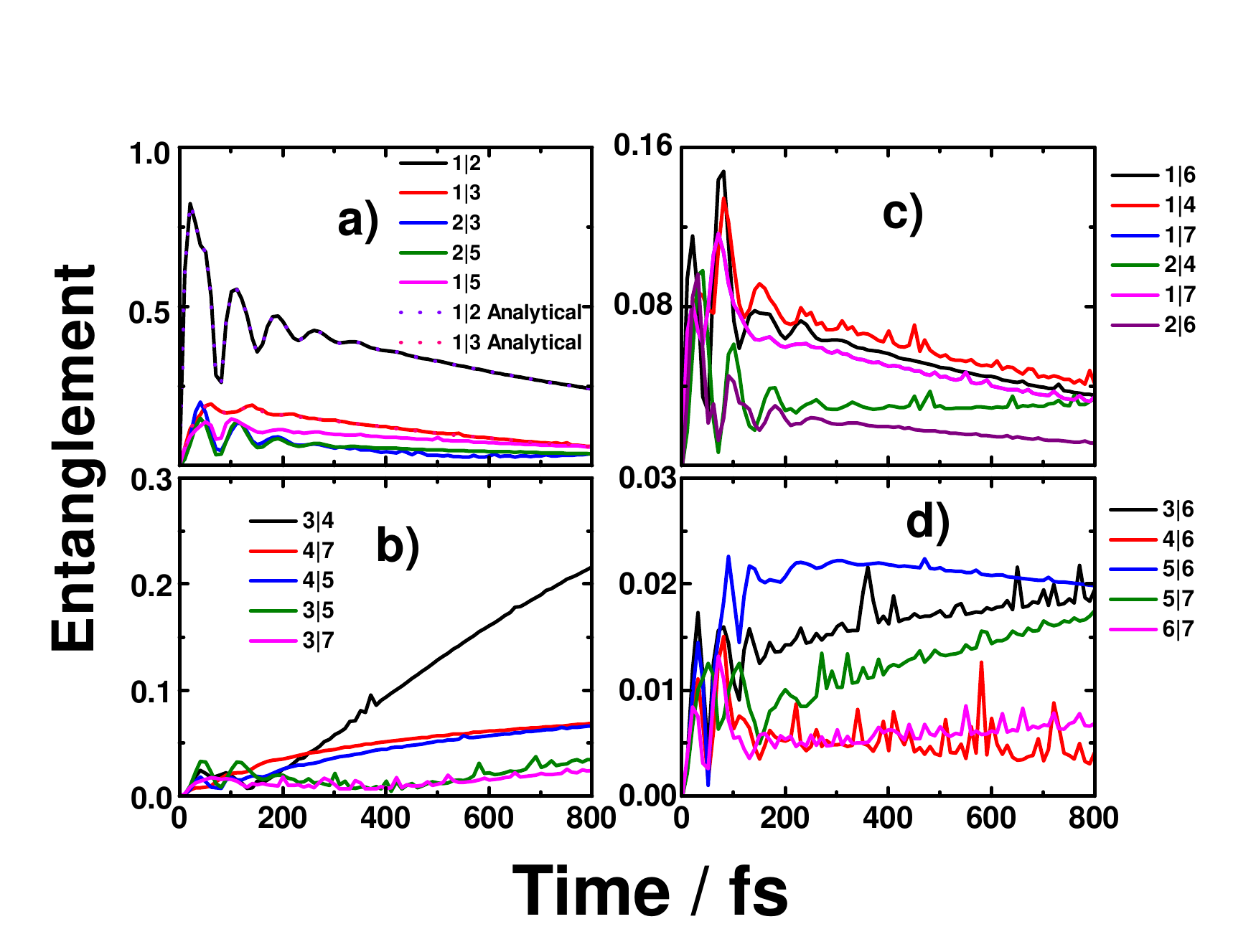}
\caption{Evolution of pairwise concurrences in the FMO complex when site one is initially excited at $T=77\unit{K}$.  This Figure shows all $21$ pairwise concurrences computed by the convex roof - these are equal to $\sqrt{\eta_S}$ for each subsystem of two sites, computed across the single bipartition of the pair. For entanglements $1|2$ and $1|3$ we also plot the exact concurrence - the agreement is good enough that the difference between the convex roof and the exact calculation is not visible. Because the monogamy bound is saturated in the single exciton manifold, these $21$ measures determine the entanglement of any single site with any subset of the others.}
\label{PairwiseS1}
\end{center}
\end{figure}

\begin{figure}
\begin{center}
\includegraphics[width=0.75\textwidth]{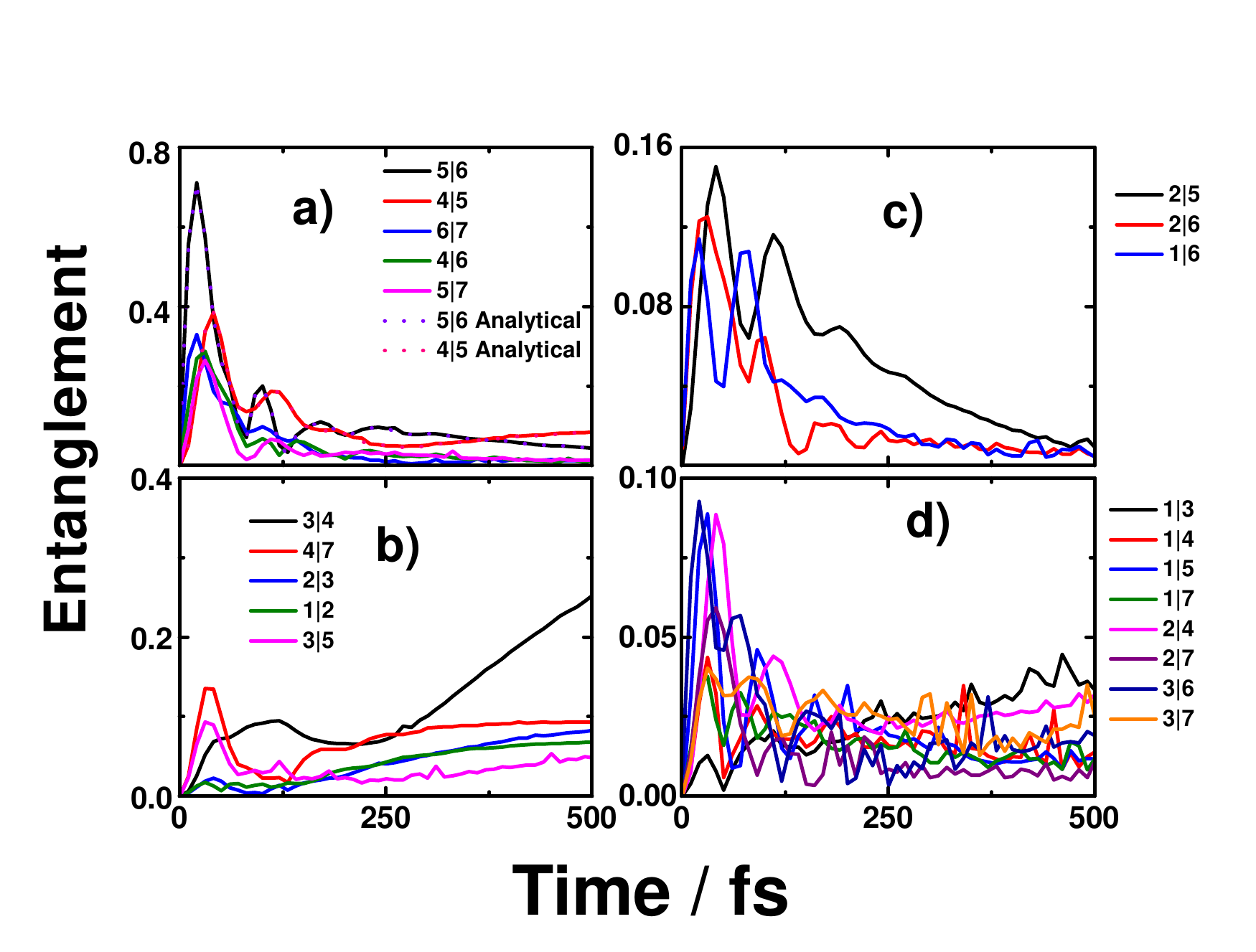}
\caption{Entanglement evolution in the FMO complex when site six is initially excited at  $T=77\unit{K}$.  This Figure shows all $21$ pairwise entanglements computed by the convex roof. For entanglements $5|6$ and $4|5$ we also plot the exact concurrence - the agreement is good enough that the difference between the convex roof and the exact calculation is not visible.}
\label{PairwiseS2}
\end{center}
\end{figure}

In Figure~\ref{PairwiseS1} we plot the entanglement evolution of the FMO complex when site $1$ is initially excited at $T=77\unit{K}$.  Figure~\ref{PairwiseS1} shows all 21 pairwise concurrences computed by the convex roof. For entanglements $1|2$ and $1|3$ we also plot the exact concurrence - the agreement is good enough that the difference between the convex roof and the exact calculation is not visible. In Figure~\ref{PairwiseS2} we plot the same data when site $6$ is initially excited at $T=77\unit{K}$. For bipartitions $5|6$ and $4|5$ we also plot the exact concurrence - again the agreement is good enough that the difference between the convex roof and the exact calculation is not visible. Figures~\ref{PairwiseS1} and~\ref{PairwiseS2} show the ordering $1|2>1|3>1|5$ as the significant concurrences for site one initially excited and $5|6>4|5$  as the significant concurrences when site six is initially excited. These results are consistent with those of~\cite{Sarovar:2010p6945}.

These results on two chromophore subsystems help us identify a pathway involving sites $1234$ as significant for exciton transport when site $1$ is initially excited, and a pathway involving sites $6543$ as significant for exciton transport when site $6$ is initially excited. This is consistent with prior results on pairwise entanglement~\cite{Sarovar:2010p6945,Whaley:2010p8611}. These results also validate our convex roof computations, at least for the case of two chromophore systems. It is perhaps unsuprising that the convex roof optimization performs well in that setting and so we now turn our attention to larger subsystems.

\subsection{Three site subsystems}

\begin{figure}
\begin{center}
\includegraphics[width=0.75\textwidth]{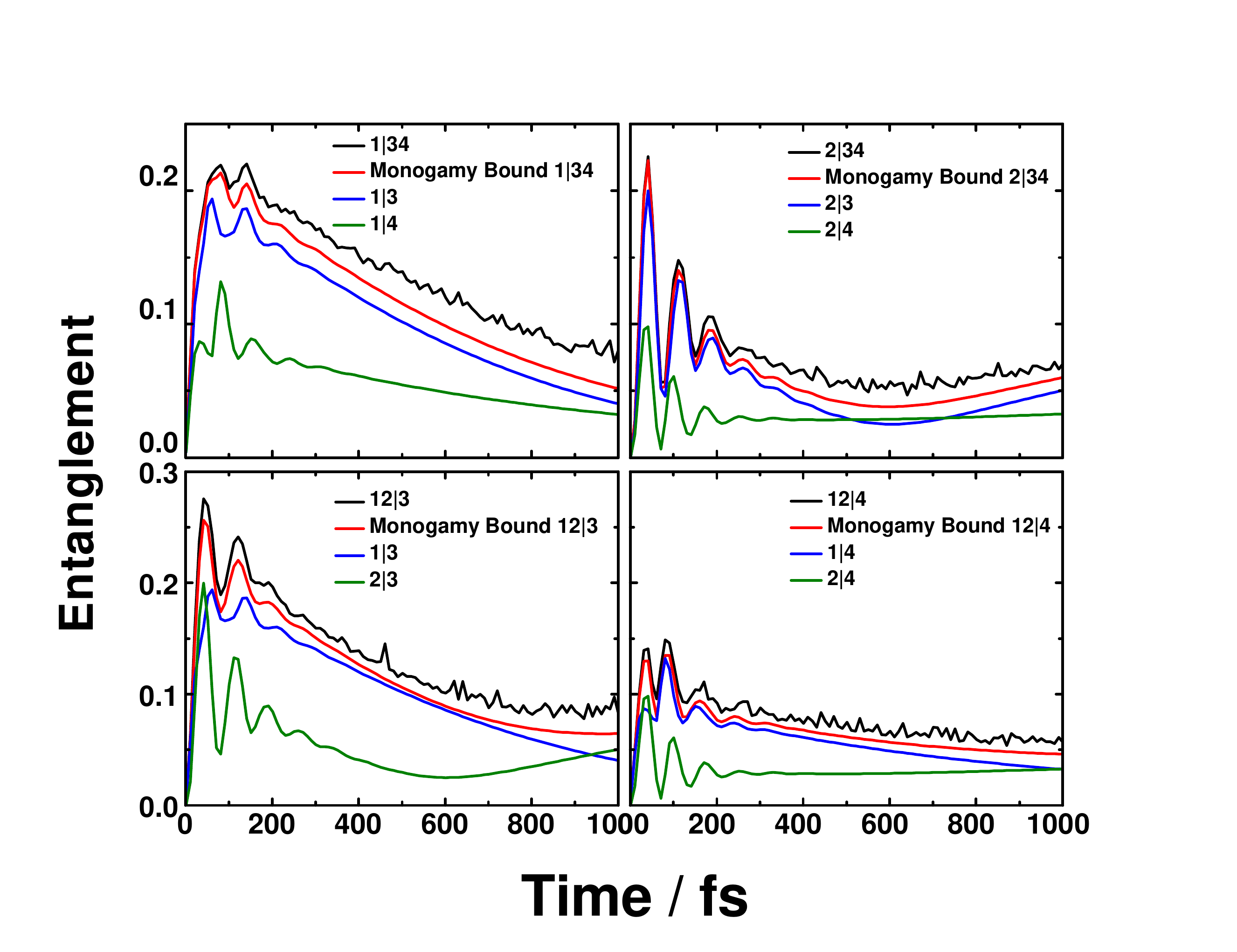}
\caption{Monogamy bound and convex roof computation of entanglements $1|34$, $2|34$,  $12|3$ and $12|4$. Particularly in the first $200~{\rm fs}$ the convex roof closely matches the monogamy bound.}
\label{monogamy}
\end{center}
\end{figure}

For any triplet of chromophores there are three bipartitions (for example, $1|23$, $2|13$ and $3|12$). Figure~\ref{monogamy} shows results for subsystems of three chromophores. We compute the entanglement measures $\sqrt{\eta_S}$ using the convex roof procedure among the triples of chromophores $134$ (for bipartition $1|34$, $S=1$), $234$ (for bipartition $2|34$, $S=2$) , $123$ (for bipartition $12|3$, $S=3$), and $124$ (for bipartition $12|4$, $S=4$). We also compute these same entanglements from the pairwise entanglements computed in the previous section using the monogamy bound.  The results shown in Figure~\ref{monogamy} illustrate the utility of the monogamy bound~\cite{SanKim:2008p8940} as a method of evaluating performance of the convex roof optimization. The convex roof performs well for three qubits, closely matching the monogamy bound.

\begin{figure}
\begin{centering}
\includegraphics[width=0.75\textwidth]{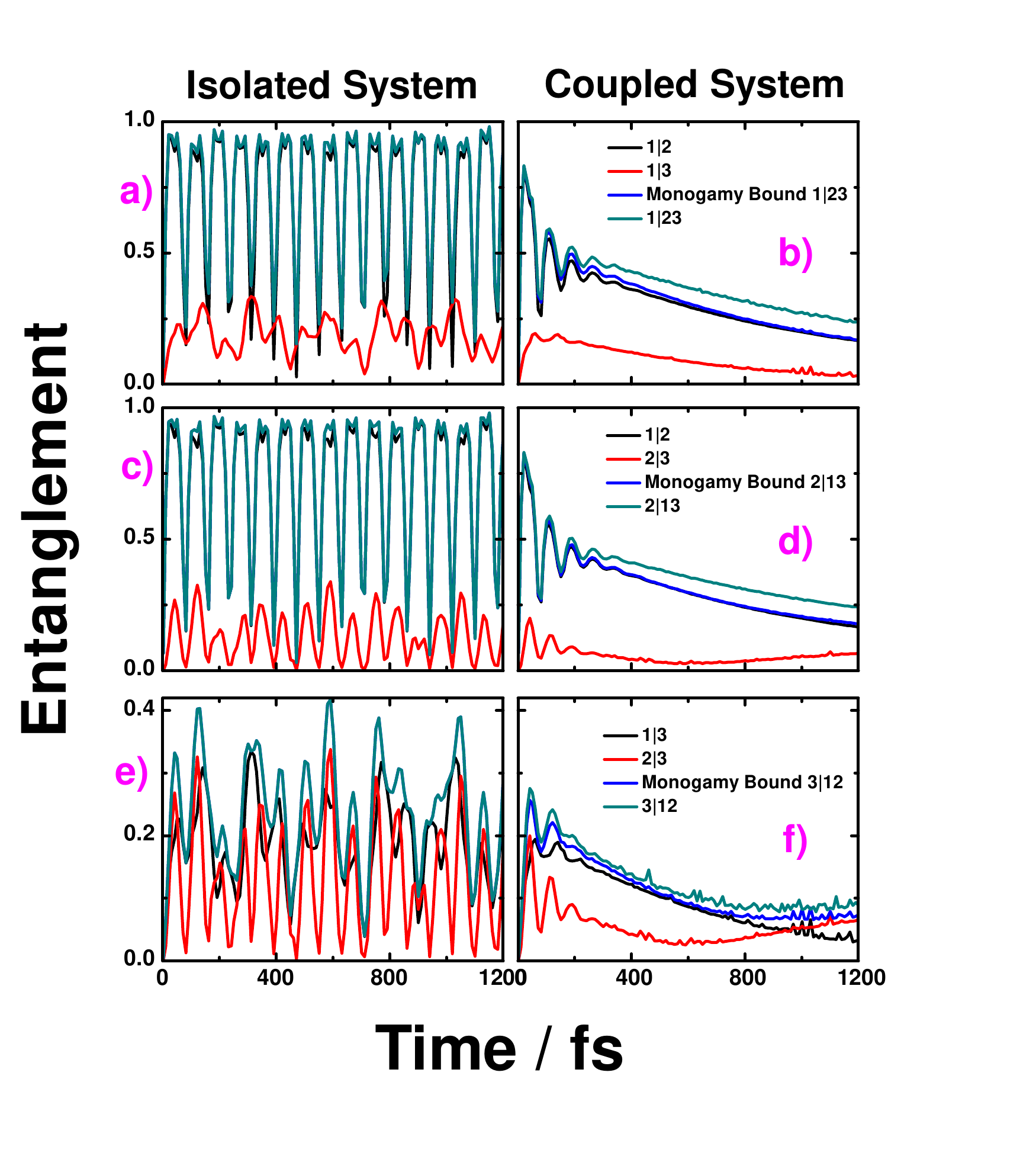}
\par\end{centering}
\caption{Entanglement evolution of FMO complex when site $1$ is initially
excited at cryogenic temperature $T=77\unit{K}$. The triplet site
entanglement among site $1$, $2$ and $3$ and also the pairwise
site entanglement between any two of site $1$, $2$ and $3$ are
plotted. The left panel shows the dynamics of the entanglement for
the system alone while the right considers the effect of the environment
\label{S1pathway}}
\end{figure}

Figure \ref{S1pathway}  shows the evolution of entanglement measures $\sqrt{\eta_S}$ across bipartitions $1|23$, $2|13$ and $3|12$ among the triplet of sites $123$ in both the isolated system and the system coupled to the environment. The left side of Figure \ref{S1pathway} shows the evolution of entanglement for the isolated system, while the right side are results from the scaled HEOM approach. For the isolated system, the oscillations in population and entanglement will last forever. By comparison with the open system case, it is obvious that the environment has the effect of eliminating the coherent oscillations characteristic of closed system quantum dynamics. Both the isolated and the system with environment hit the maximum and minimum values at the same time during the evolution, which shows that the oscillations in the open system case are indeed the remnants of the coherent behavior in the closed system case. The entanglement evolution is not as smooth as ref. \cite{Sarovar:2010p6945}, because the simulation data has been sampled every $10\unit{fs}$ in order to perform the entanglement calculations.


Fig.\ref{S1pathway}b shows  $\sqrt{\eta_1}$, the entanglement of subsystem $123$ across partition $1|23$. The pairwise concurrences across bipartitions $1|2$ of subsystem $12$ and $1|3$ of subsystem $13$ and the monogamy bound is also shown. The time series of $\sqrt{\eta_1}$  across bipartition $1|23$ reflects the coherent oscillation of the population and the time over which these oscillations last  is the same as that in the population evolution which is around $400\unit{fs}$. The entanglement $\sqrt{\eta_S}$ across bipartition $1|23$ is predominantly due to the pairwise entanglement evolution between site $1|2$, particularly during the first few oscillations ($t<200\unit{fs}$). Beyond $200\unit{fs}$, the value of the measure $\sqrt{\eta_1}$, the entanglement of subsystem $123$ across partition $1|23$, becomes slightly larger than the pairwise entanglement site $1|2$, indicating that sites one and three have become entangled at this time.

Fig. \ref{S1pathway}d, shows $\sqrt{\eta_2}$, the entanglement of subsystem $123$ across the bipartition $2|13$. This time series is similar to that of  $1|23$, again because $\sqrt{\eta_2}$ is dominated by the entanglement of sites $1$ and $2$. Another interesting phenomena is the pairwise concurrence across bipartition $2|3$, which also shows coherent oscillations. Although the value of the concurrence is much smaller compared with the entanglement between site $1|2$, the oscillations of $2|3$ share the same frequency and hit the maximum and minimum value simultaneously.

Fig. \ref{S1pathway}f shows $\sqrt{\eta_3}$,  the entanglement of the triplet $123$ across the partition $3|12$, which is much smaller than the entanglement across bipartitions $1|23$ and $2|13$ and does not show significant coherent oscillations. For this case, in which site $1$ is initially excited, the dominant pairwise entanglement is $1|2$, which is consistent with the other results in the literature~\cite{Sarovar:2010p6945,Caruso:2010p8915,Caruso:2009p8916}. Hence, one may understand the smaller value of this measure of entanglement by noticing that it is computed across a bipartition that does not separate sites $1$ and $2$.

As a result, we conclude that in this pathway: during the coherent evolution period (first $200\unit{fs}$), sites $3$ and $4$ are competing with each other to be entangled with sites $1$ and $2$. However, when the coherent evolution disappears, the entanglement between site $3$ and $4$ becomes dominant.

\begin{figure}
\begin{centering}
\includegraphics[width=0.75\textwidth]{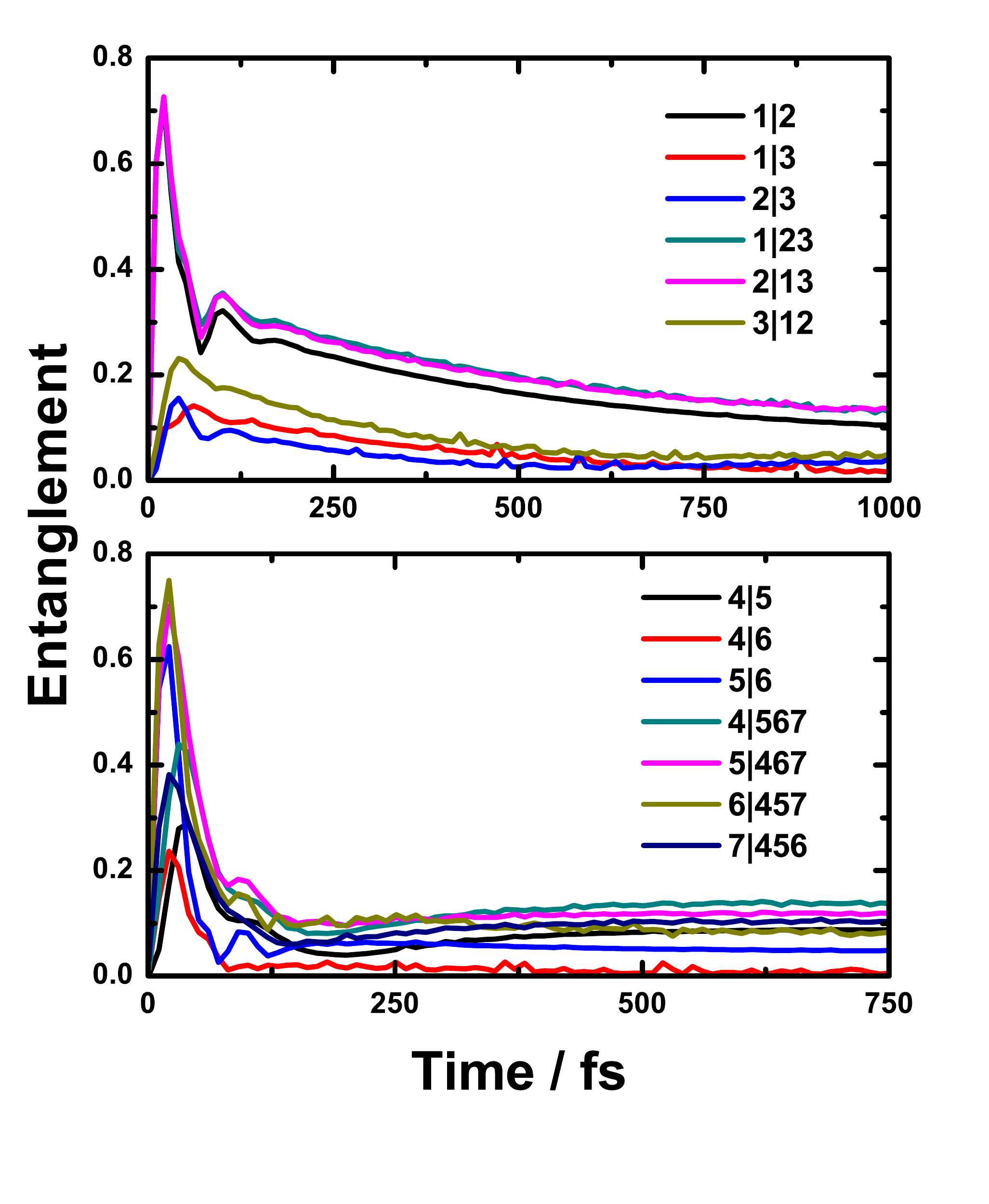}
\par\end{centering}

\caption{Time evolution of entanglement for multiple sites at $T=300\unit{K}$. In the upper panel the entanglement measures $\sqrt{\eta_S}$ across the indicated bipartitions among sites $1$, $2$ and $3$ are shown when site $1$ is initially excited. For the lower panel, site $6$ is initially excited. \label{300KEn}}
\end{figure}

In order to check the effect of temperature on the entanglement evolution, we plotted the entanglement evolution at room temperature ($T=300\unit{T}$) for both site $1$ and site $6$ initially excited. The results at $300K$ are shown in Figure~\ref{300KEn}. By comparing with the evolution at $T=77\unit{K}$ shown in Figure~\ref{PairwiseS1}, the coherent oscillations were reduced from $4$ to $2$ oscillations and the length of coherent oscillations was also reduced from $400\unit{fs}$ to $<250\unit{fs}$. The maximum entanglement during the evolution was also reduced due to the increase in temperature. For example, the maximumvalue of the masures $\sqrt{\eta_S}$ for bipartition $1|23$ of sites $123$ is $0.85$ at $77\unit{K}$ while that is around $0.73$ when $T=300\unit{K}$. In addition the entanglement goes to the equilibrium state much faster at $300\unit{K}$ than at $T=77\unit{K}$. It takes around $7\unit{ps}$ for the system to arrive at the equilibrium state at $T=77\unit{K}$, while at $T=300\unit{K}$ this takes around $1.5\unit{ps}$. These results all confirm that the scaled HEOM approach correctly reproduces the known effects of increasing temperature on the evolution of entanglement.

\subsection{Four site subsystems}

There are four distinct bipartitions of the system into one site plus the rest and we may use the monogamy bounds to evaluate the performance of our convex roof calculations. However, there are also three distinct bipartitions of the four site subsystems into pairs of sites and we also compute measures $\sqrt{\eta_S}$ across these bipartitions ($12|34$, $13|24$, $14|23$).

\begin{figure}[h]
\begin{center}
\includegraphics[width=0.75\textwidth]{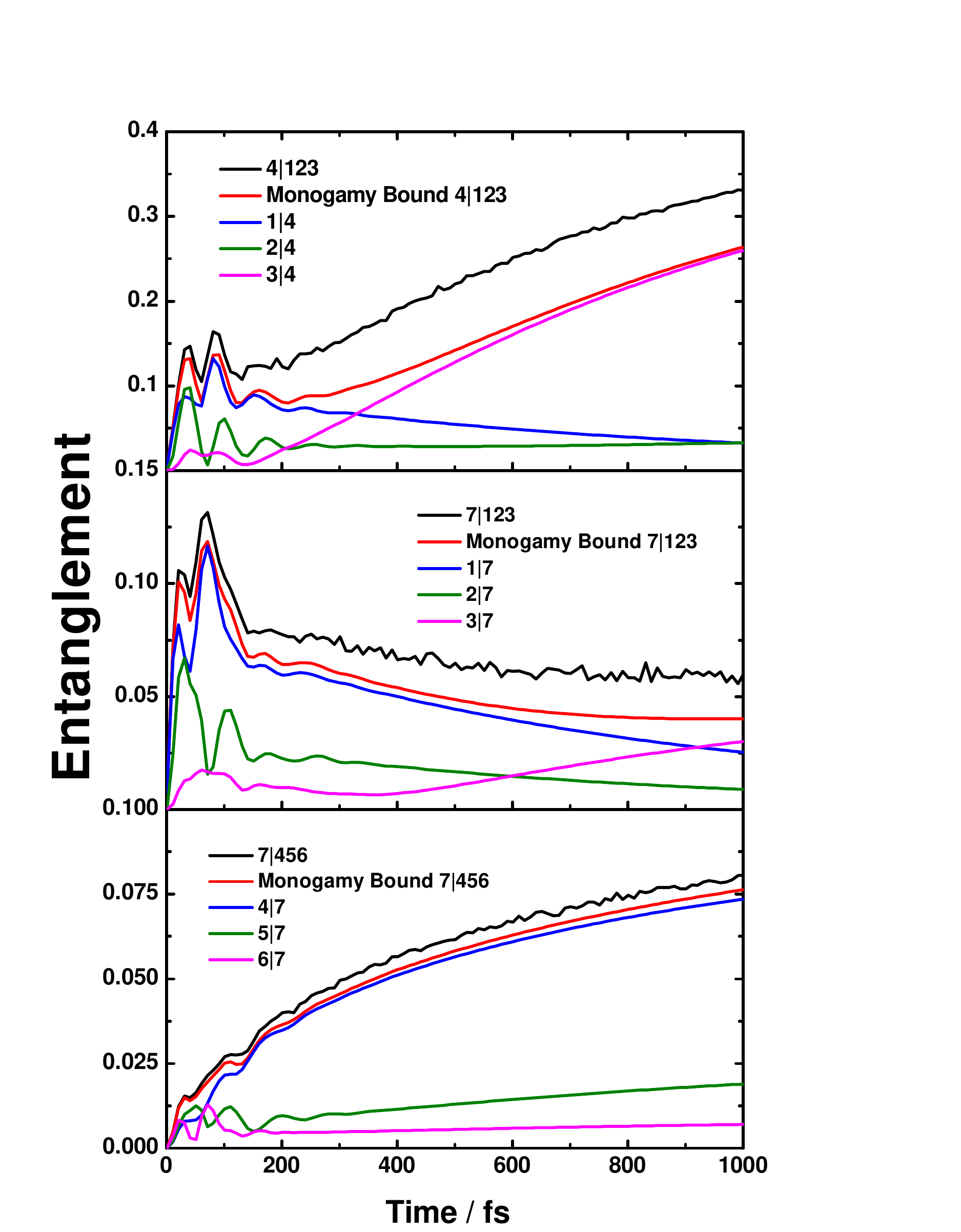}
\caption{Measures of entanglement and monogamy bounds in a four qubit system when site $1$ is initially excited at temperature $T=77\unit{K}$. The measures $\sqrt{\eta_S}$ computed for bipartitions $4|123$, $7|123$ and $7|456$ by the convex roof and together with the monogamy bound are shown here. We see a larger variation in performance of the convex roof optimization here, with a smaller difference between the upper (convex roof) and lower (monogamy) bounds for $7|456$ and $7|123$ than for $4|123$. }
\label{Quad1_3_S1}
\end{center}
\end{figure}

In Figure~\ref{Quad1_3_S1} we evaluate the performance of our convex roof optimization using the monogamy bounds. As one can see, the difference between the upper and lower bounds is larger than for two and three chromophore systems, but is significantly smaller in the case shown in the lower panel where the values of the measures $\sqrt{\eta_S}$ are rather small ($\sqrt{\eta_S}<0.1$ for $7|456$).

\begin{figure}
\begin{centering}
\includegraphics[width=0.75\textwidth]{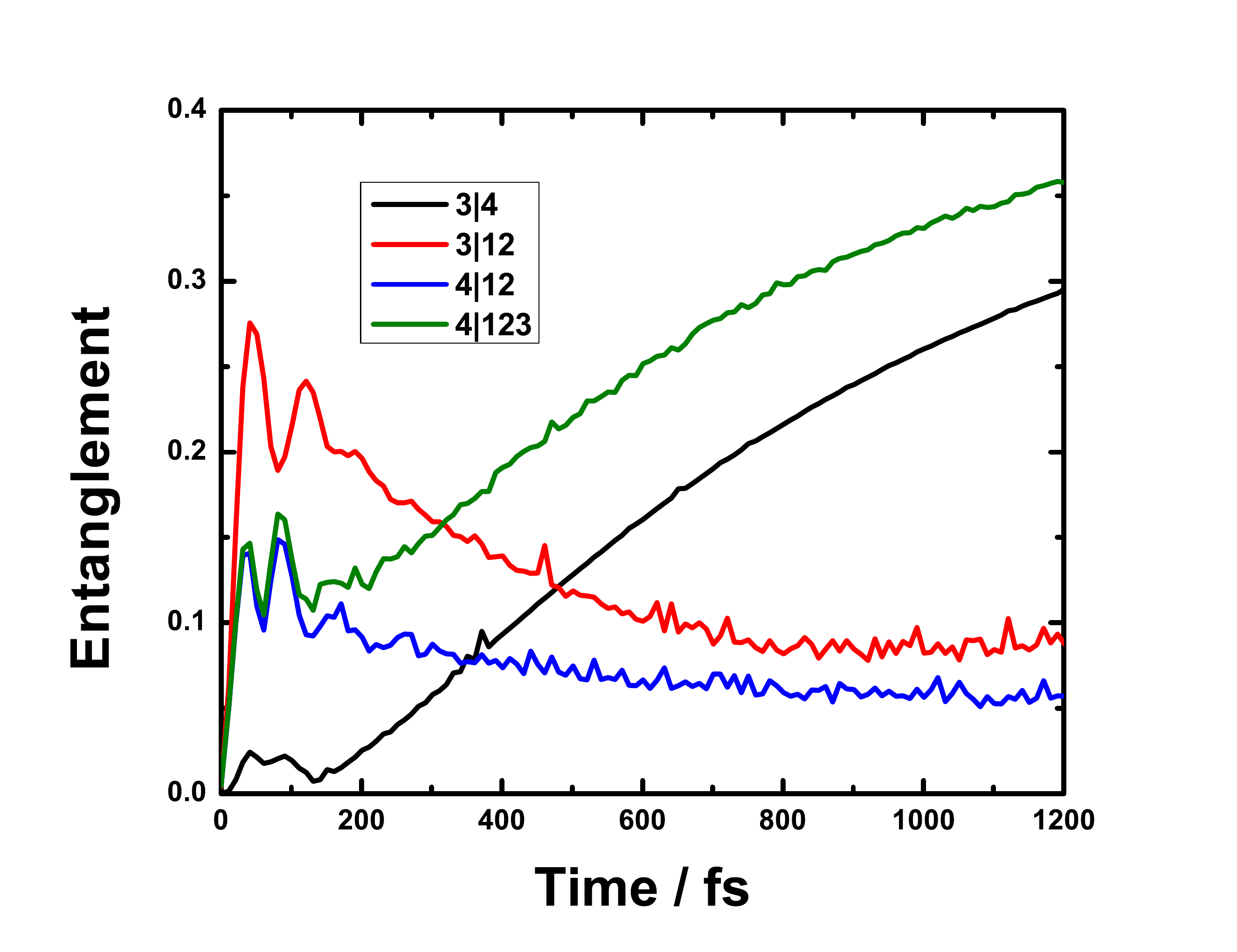}
\par\end{centering}
\caption{Time evolution of various entanglement measures for subsystem $1234$ for site $1$ initially excited at $T=77\unit{K}$. The concurrence for subsystem $34$ across bipartition $3|4$ and the measures $\sqrt{\eta_S}$ for subsystems $123$ and $124$ across bipartitions $3|12$ and $4|12$, respectively, are also shown. The measure $\sqrt{\eta_S}$ across bipartition $4|123$ is also shown. \label{S1Site34}}
\end{figure}

Next we examine the different roles of sites $3$ and $4$ in the pathway involving sites $1234$ for the case where site $1$ is initially excited. It is known that the destination of this pathway is the pair of sites  $34$. However, the detailed roles of these two sites during the entanglement evolution is still not clear.  Figure \ref{S1Site34} shows the evolution of the entanglement measure $\sqrt{\eta_S}$ for the subsystem of chromophores $1234$ across partition $4|123$. The concurrence for the pair $34$ across partition $3|4$, and the measures $\sqrt{\eta_S}$ for triplets $123$ and $124$ across partitions $3|12$ and $4|12$ are also shown for comparison. Within the first $200$ fs we see coherent oscillations in which $3|12$ and $4|12$ are in antiphase, but where $4|123$ is in phase with $4|12$. The concurrence $3|4$ evolves in lockstep with the measure $\sqrt{\eta_S}$ across bipartition $4|123$ after $200$fs. The entanglement of $3|12$ and $4|12$ are also evolving comparably after $200$fs. This behavior is suggestive of an initial period (the first $200$ fs) in which the entanglement of chromophore $4$ with $123$ is fixed by its entanglement with chromophores $12$, and then a long - time behavior in which chromophore $4$ is entangled with chromophore $3$. This is consistent with a picture of energy transport in which a delocalized exciton passes from chromophores $12$ to chromophores $34$ - eventually landing at chromophore $3$.

\begin{figure}
\begin{center}
\includegraphics[width=0.75\textwidth]{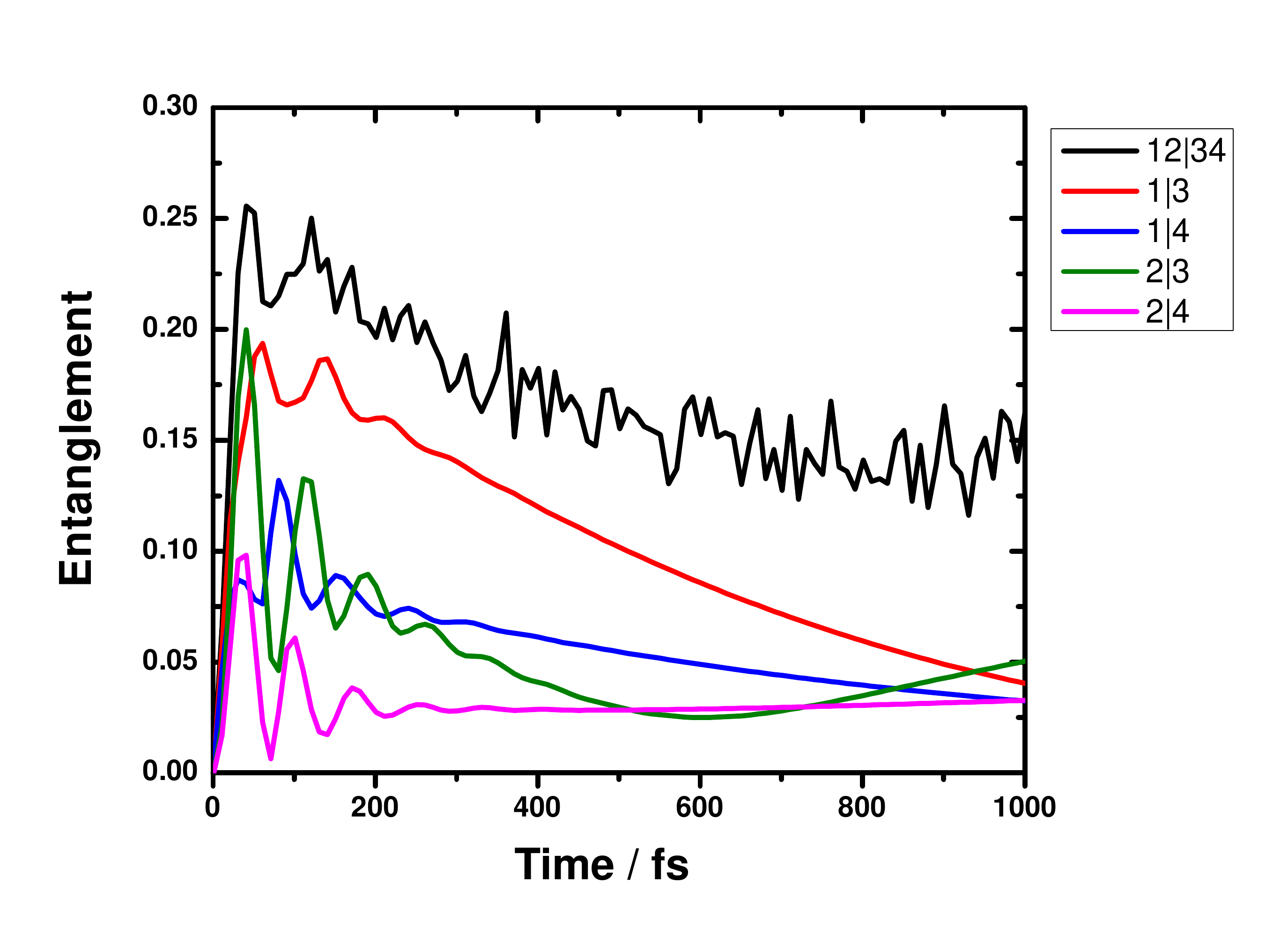}
\caption{Entanglement measures for the four chromophore subsystem $1234$ when site $1$ is initially excited at temperature $T=77\unit{K}$. The measure $\sqrt{\eta_S}$ across bipartition $12|34$ was computed via the convex roof procedure and is shown here, together with the concurrences for pairs of chromomphores $13$, $14$, $23$ and $24$.  We note that in this case, we see that the entanglement $12|34$ evolves similarly to both the $1|3$ and $2|3$ concurrences.}
\label{Quad2_2}
\end{center}
\end{figure}

In Figure~\ref{Quad2_2} we show the measure $\sqrt{\eta_S}$ for subsystem $1234$ across partition $12|34$. This tells us the entanglement between pairs of chromophores $12$ and $34$ for the case where site $1$ is initially excited. Comparison of this figure with Figure~\ref{S1Site34} is instructive, as we see that the entanglement between the pairs of chromophores $12$ and $34$ is decreasing after the first $200$ fs - following the falling value of the concurrence of the pair $13$ across bipartition $1|3$. This makes sense in a picture of transport in which $12$ are the chromophores receiving the exciton when it is injected and $34$ receive the exciton before it passes to the reaction center.

\begin{figure}
\begin{centering}
\includegraphics[width=0.75\textwidth]{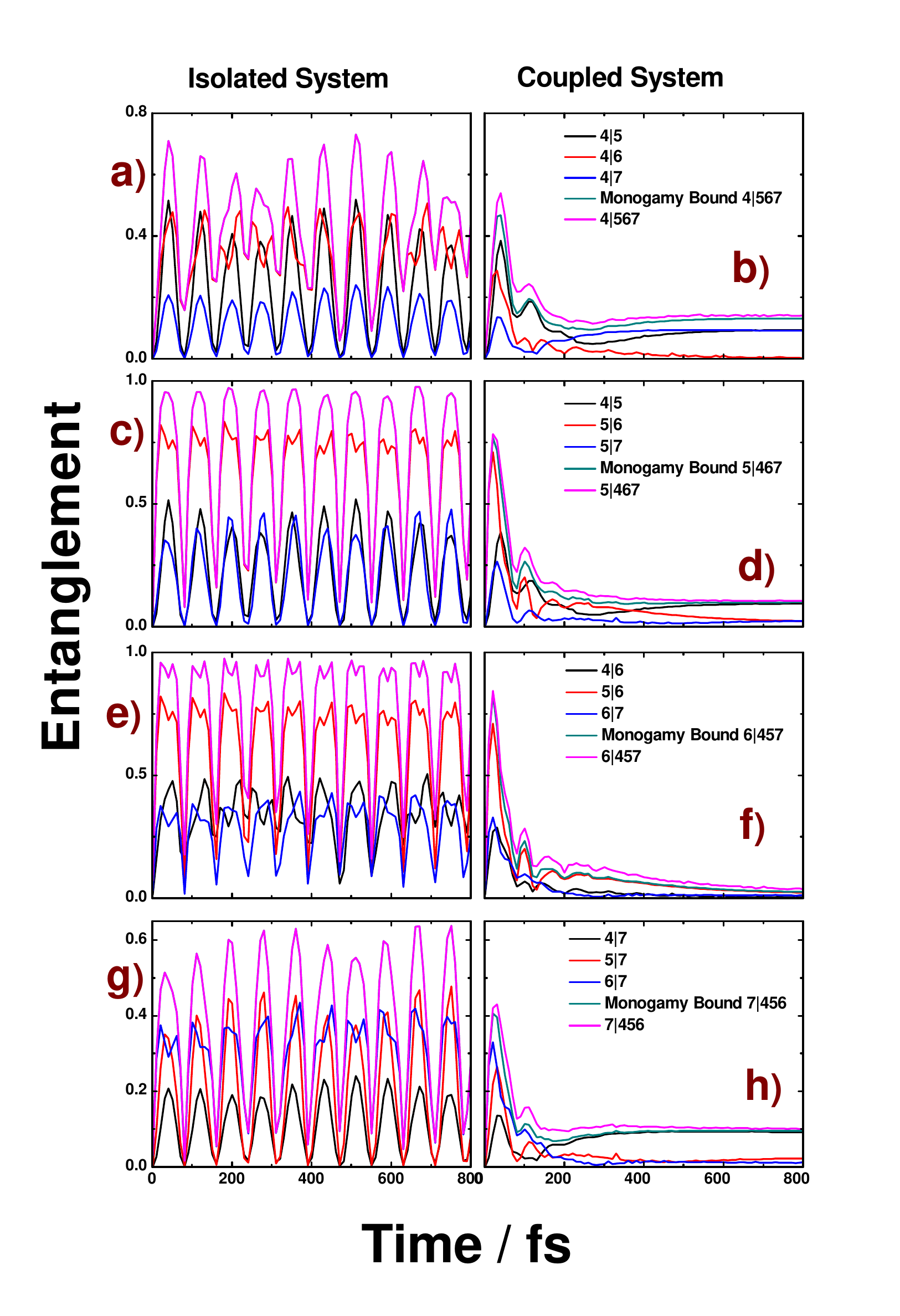}
\par\end{centering}
\caption{Time evolution of concurrences and measures $\sqrt{\eta_S}$ for the FMO complex when site $6$ is initially excited at temperature $T=77\unit{K}$. The measures $\sqrt{\eta_S}$ are shown are for subsystem $4567$ across bipartitions $4|567$, $5|467$, $6|457$ and $7|456$. We also show the concurrences among the pairs of sites that determine the concurrence bounds for the measures $\sqrt{\eta_S}$ across bipartitions $4|567$, $5|467$, $6|457$ and $7|456$, and the concurrence bounds themselves. The left panel shows the isolated system evolution and the right panel shows the open system dynamics with environment. \label{S6pathway}}
\end{figure}

We now turn to the case in which site $6$ is initially excited. Fig. \ref{S6pathway} shows the evolution of entanglement measures $\sqrt{\eta_S}$ for the subsystem $4567$ in both the isolated and open system case. Similar to the case where site $1$ is initially excited, the measures $\sqrt{\eta_S}$ display coherent oscillations which persist as long as the oscillations in the population. The most significant concurrence is that for subsystem $56$ across bipartition $5|6$, for which the maximum value is $0.8$. The second most important pairs are sites $4|5$ and $4|6$, which have the maximum concurrence around $0.4$. On the other hand, the coherent oscillations for all three pairs share the same frequency and evolution trend after the $1$st beating. For subsystems $4567$ the measures $\sqrt{\eta_S}$ across bipartitions $6|457$ and $5|467$ have similar amplitude and time evolution. However, the measures $\sqrt{\eta_S}$  $3|567$ and $4|567$ are much smaller compared with the above two. Comparison of the measures $\sqrt{\eta_S}$ computed across bipartitions $4|567$, $5|467$, $6|457$ and $7|456$ by the convex roof (which gives an upper bound) with the monogamy bounds (which are lower bounds) shows that the convex roof is performing well in this case.

\subsection{Five site subsystems}

For five qubits subsystems there are five partitions of the subsystem that divide one site from the other four, and ten partitions that divide two sites from the other three. We proceed as for the four site system, using the monogamy relations to evaluate the performance of the convex roof measure.

\begin{figure}[h]
\begin{centering}
\includegraphics[width=0.75\textwidth]{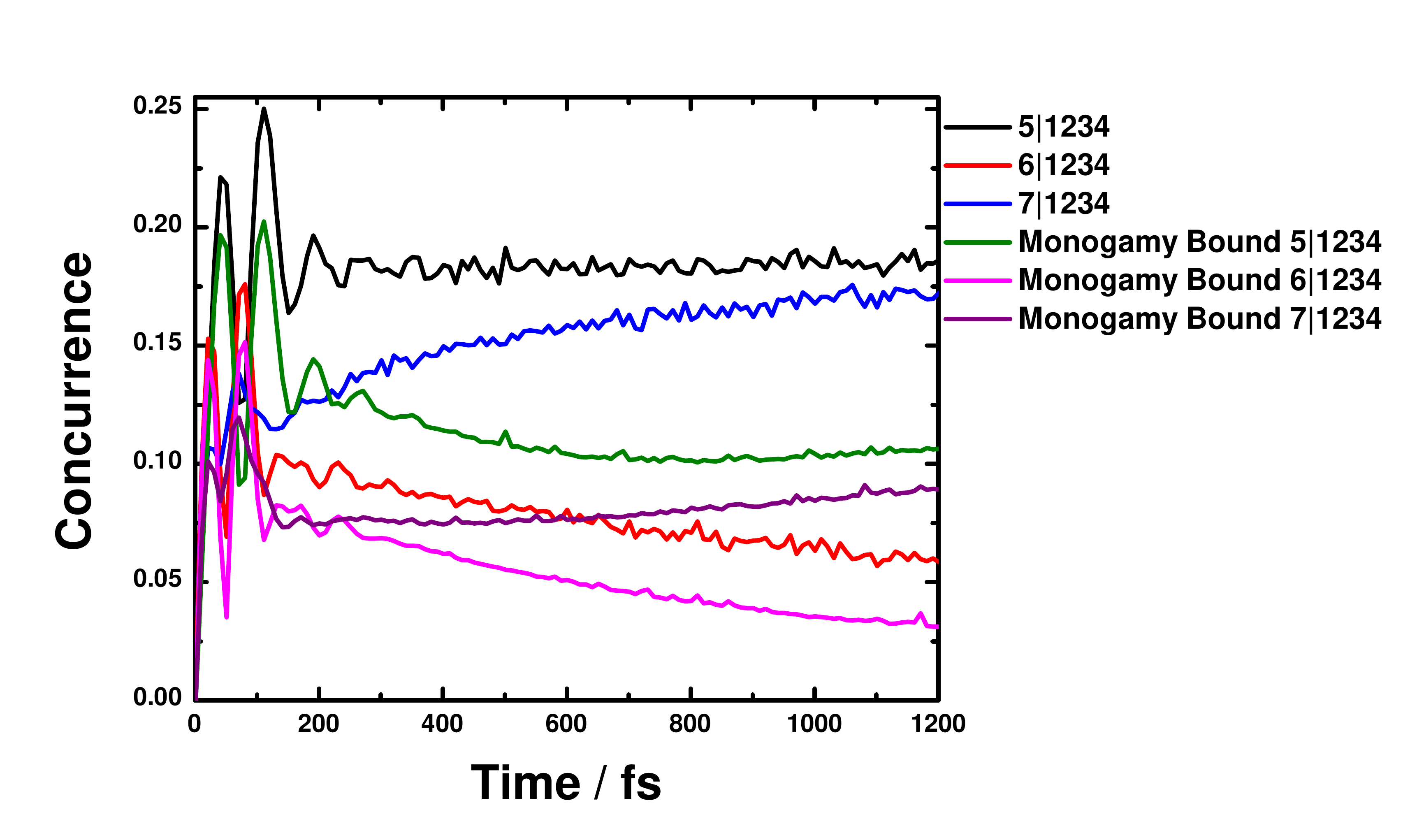}
\par\end{centering}
\caption{Evolution of entanglement measures $\sqrt{\eta_S}$ for the subsystems of chromophores $12345$, $12346$ and $12347$ in the FMO complex  at cryogenic temperature $T=77\unit{K}$ for site $1$ is initially excited. The measures $\sqrt{\eta_S}$ are computed across bipartitions $5|1234$, $6|1234$ and $7|1234$ and the corresponding monogamy bounds are also shown.  Site $1$, $2$, $3$ and $4$ are sites evolved in the population pathway under this initial condition, and this data indicates that the entanglement of this subset ($1234$) of chromophores with the other three chromophores is small. \label{OffpathS1}}
\end{figure}

Fig. \ref{OffpathS1} shows the evolution of entanglement measures $\sqrt{\eta_S}$ for subsystem $12345$. The measures $\sqrt{\eta_S}$ across the three bipartitions $5|1234$, $6|1234$ and $7|1234$ are all small  ($<0.25$) during the full time evolution. This shows that when site $1$ is initially excited, the measures $\sqrt{\eta_S}$ are only large between sites in the pathway, which are sites $1$, $2$, $3$ and $4$.  We also plotted the monogamy bounds in Fig. \ref{OffpathS1},  this shows that, unsurprisingly, the difference between the convex roof optimization and the monogamy bound is larger in this case - likely showing that the convex roof optimization is not performing as well in the five qubit case as it does for three and four qubits.

\begin{figure}
\begin{centering}
\includegraphics[width=0.75\textwidth]{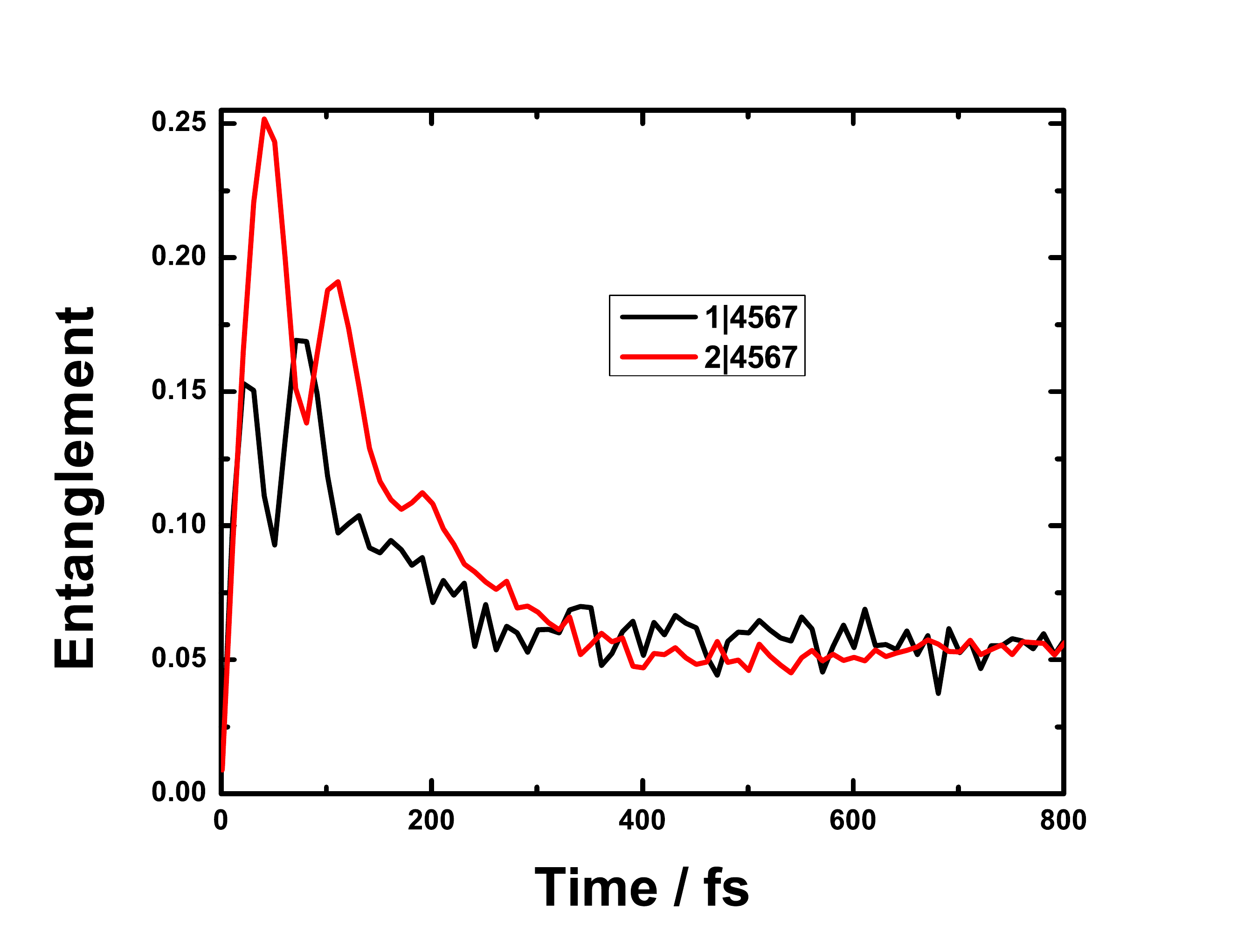}
\par\end{centering}
\caption{Time evolution of entanglement measures $\sqrt{\eta_S}$ in the FMO complex for site $6$ initially excited at cryogenic temperature $77\unit{K}$. The measures $\sqrt{\eta_S}$ are shown for subsystem $14567$ across bipartition $1|4567$ and subsystem $24567$ across bipartition $2|4567$. \label{OffpathS6}}
\end{figure}

For the case in which site $1$ is initially excited, we only see significant values of the entanglement measures within the sites $1234$ in the pathway. We would like to know if this is also the case when site $6$ is initially excited. Figure \ref{OffpathS6} shows the entanglement measure $\sqrt{\eta_S}$ for subsystems $14567$ across bipartition $1|4567$ and subsystem $24567$ across bipartition $2|4567$.  The maximum value of these entanglement measures is around $0.25$, which is much smaller compared than that for measures computed across bipartitions of the  subsystem $4567$. This is consistent with the idea that entanglement is concentrated among the sites evolved in a specific pathway, with different pathways for different initial conditions.

\begin{figure}
\begin{centering}
\includegraphics[width=0.75\textwidth]{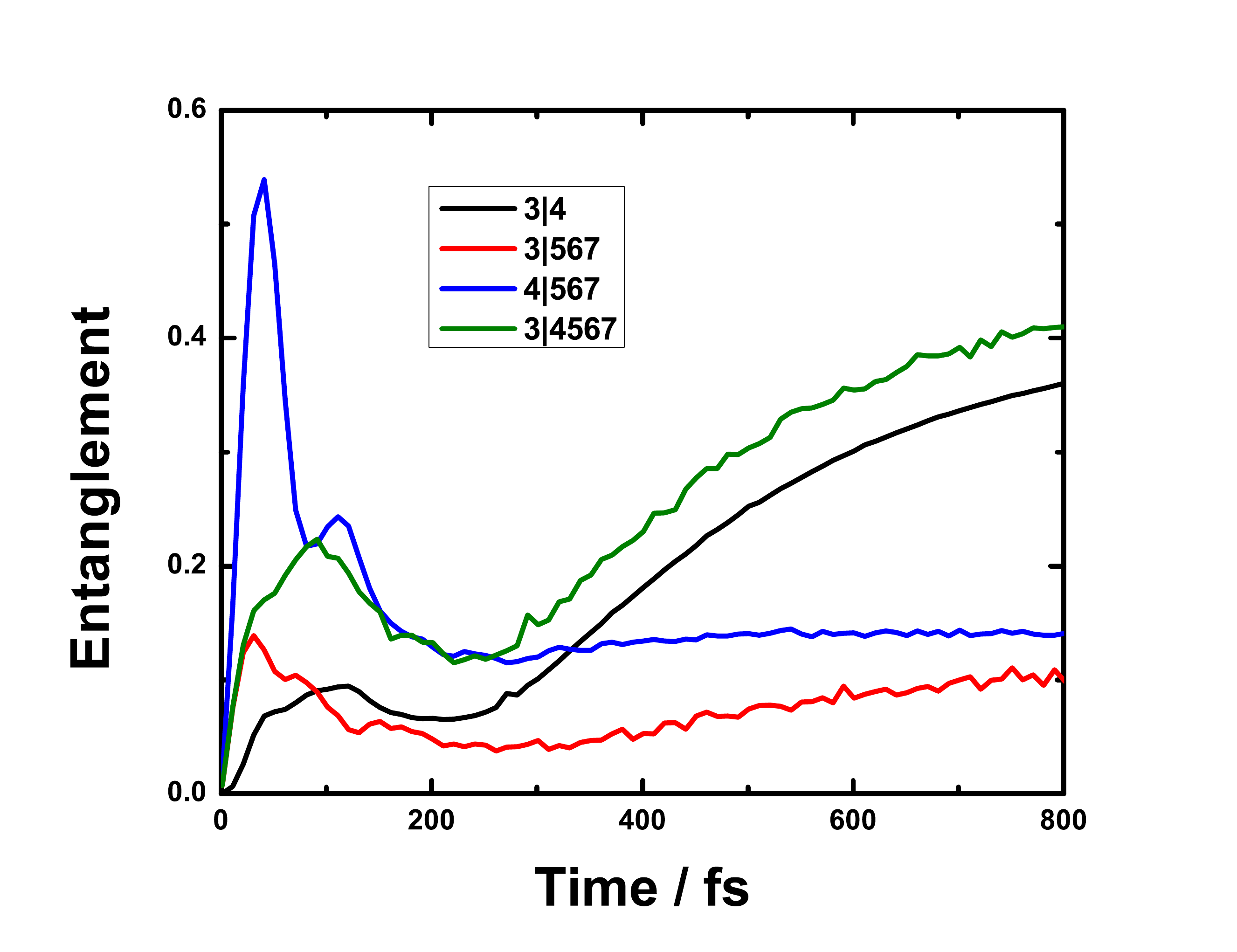}
\par\end{centering}

\caption{Time evolution of entanglement measures $\sqrt{\eta_S}$ in the FMO complex for site $6$ initially excited at $77\unit{K}$. (c.f. Figure \ref{S1Site34}). The measures $\sqrt{\eta_S}$ are computed for subsystems $3567$ across bipartition $3|567$, subsystem $4567$ across bipartition $4|567$ and subsystem $34567$ across bipartition $3|4567$.  \label{S6Site34}}
\end{figure}

As for the case when site $1$ is initially excited, we also examined the roles of sites $3$ and $4$ in the case when site $6$ is initially excited (Fig. \ref{S6Site34}). Just as in the case where site $1$ was initially excited (Figure \ref{S1Site34}) we see an initial period with coherent oscillations in the entanglement in which the entanglement of $3$ with the rest and $4$ with the rest are in antiphase. This is followed by a later period in which sites $3$ and $4$ become entangled and the entanglement of $3$ with $4567$ is dominated by the entanglement of $3$ and $4$. As a result, the dominant pairwise entanglement changes from site $5|6$ to pair $3|4$ during the transport of the exciton from the injection site at site $6$ to the final state in which it is concentrated on sites $3$ and $4$.

\subsection{Seven Site Calculations}

The nature of multipartite entanglement in the FMO complex is encoded in the bipartite entanglement across multiple partitions. In the preceding sections we have attempted to build up a picture of multipartite entanglement by considering entanglement within subsystems, and across multiple bipartitions of many subsystems. However, the performance of the convex roof optimization worsens as one moves from three to four to five site subsystems, and these optimizations are not feasible using a general treatment of the full seven site system. However, we can restrict our optimization to include ensembles constructed only within the one-exciton subspace, and by doing so calculations of the full seven chromophore system become tractable.

In this subsection we present calculations of measures $\sqrt{\eta_S}$ for all bipartitions of the full seven chromophore system. There are $63$ such bipartitions, seven of which are partitions into one chromophore plus the rest. There are $21$ distinct partitions of the FMO complex into a pair of sites and a quintuplet, and $35$ partitions of the FMO complex into a triple and a quadruple of sites.

\begin{figure}
\begin{centering}
\includegraphics[width=0.75\textwidth]{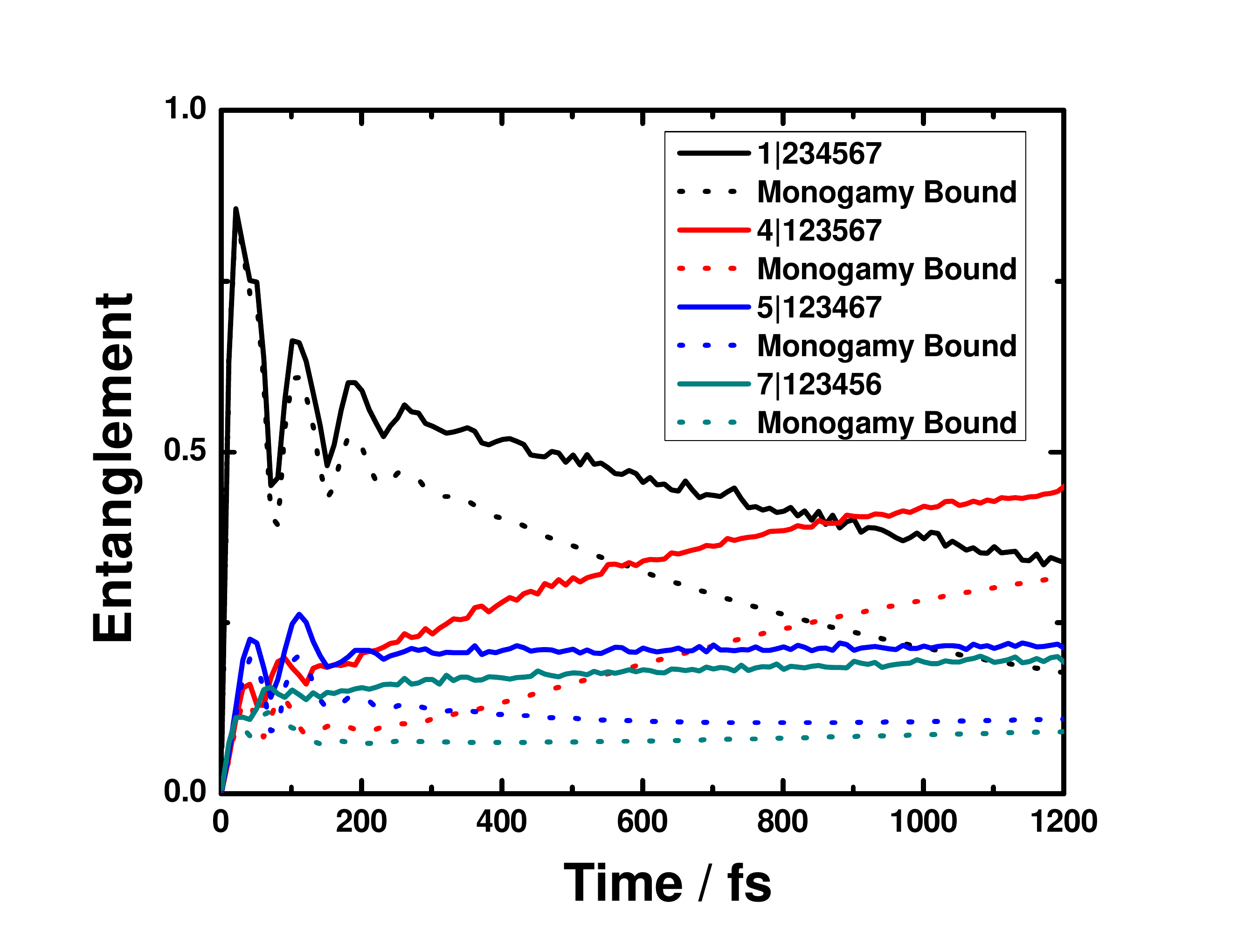}
\par\end{centering}
\caption{Entanglement measures $\sqrt{\eta_S}$ for the full FMO system at $77K$ with site one initially excited. The measures $\sqrt{\eta_S}$ are shown for the partitions $1|23456$, $4|123567$, $5|123467$, $7|123456$ (solid lines), together with the corresponding monogamy bounds (dotted lines). These results illustrate the performance of the convex roof optimization and also show that the largest of these measures is that which gives the entanglement of chromophore $1$ with the rest, $1|23456$.  \label{bounds1}}
\end{figure}

\begin{figure}
\begin{centering}
\includegraphics[width=0.75\textwidth]{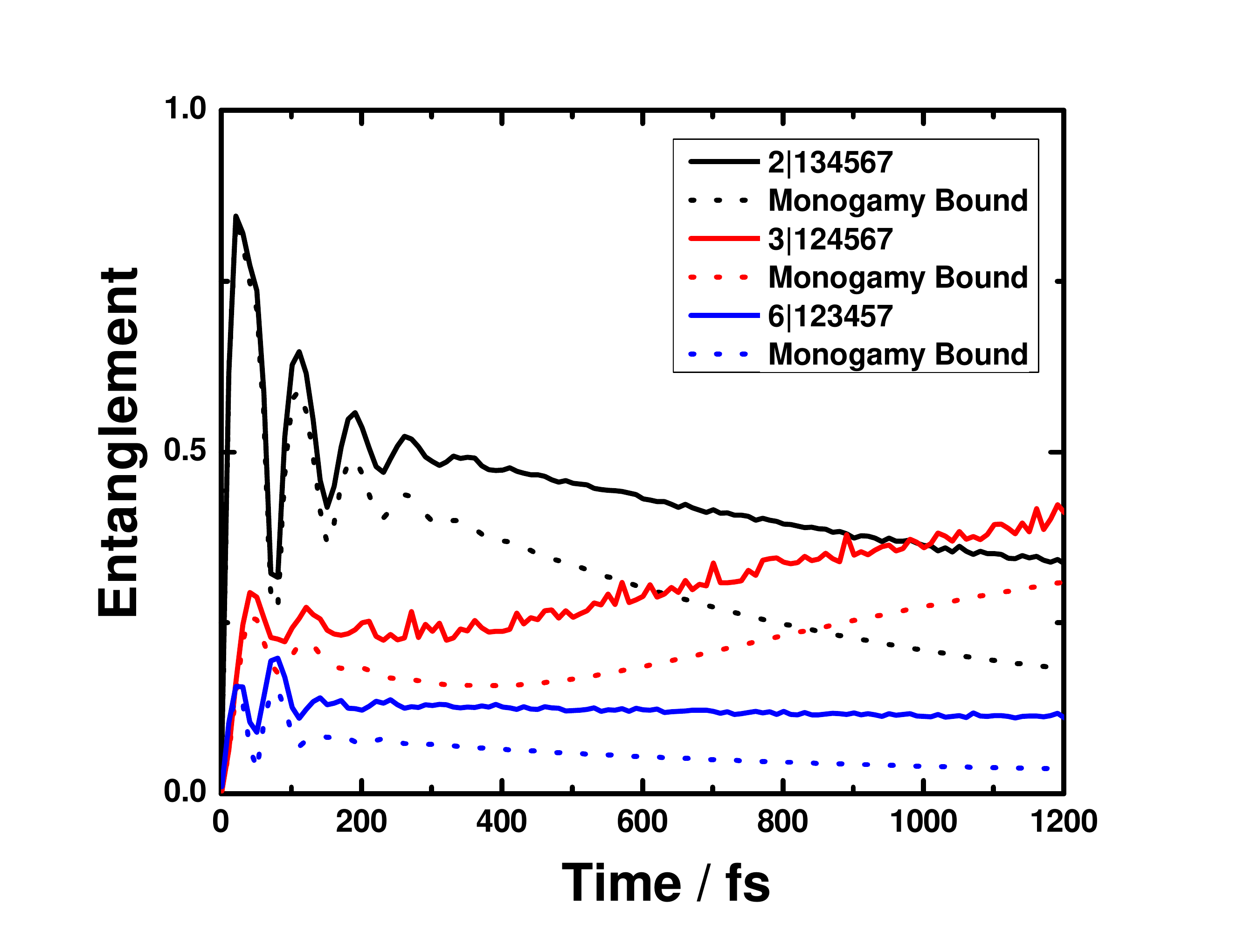}
\par\end{centering}
\caption{Entanglement measures $\sqrt{\eta_S}$ for the full FMO system at $77K$ with site one initially excited. The measures $\sqrt{\eta_S}$ are shown for the partitions $2|13456$, $3|124567$, $6|123457$ (solid lines), together with the corresponding monogamy bounds (dotted lines). These results illustrate the performance of the convex roof optimization and also show that the largest of these measures is that which gives the entanglement of chromophore $2$ with the rest, $1|23456$.\label{bounds2}}
\end{figure}

Figures~\ref{bounds1} and~\ref{bounds2} show the time evolution of all measures $\sqrt{\eta_S}$ for the seven site system across the seven bipartitions into one chromophore and the other six,  at $77K$ with site one initially excited. We compute these measures by the convex roof optimization restricted to the single exciton subspace, and also calculate the monogamy bounds. In this data we can see that only sites one and two exhbit significant ($>0.5$) values of the entanglement measures that undergo coherent oscillations. The remaining measures exhibit a rapid rise, but remain well below $0.5$ for the entire evolution.

From Figure~\ref{PairwiseS1} we see that the single pairwise concurrence of subsystem $12$ across bipartition $1|2$ exhibits coherent oscillations and large entanglement. Hence the picture of entanglement we obtain from Figures~\ref{PairwiseS1},~\ref{bounds1} and~\ref{bounds2} is that the entanglement of chromophores one and two with the rest is determined mainly by the entanglement of chromophore one with chromophore two. This is consistent with the picture obtained by examining small subsystems of the FMO complex - in which chromophores one and two initially share the excition before it moves into the other chromophores in the pathway $1234$ for the case in which chromophore one is initially excited.

\begin{figure}
\begin{centering}
\includegraphics[width=0.75\textwidth]{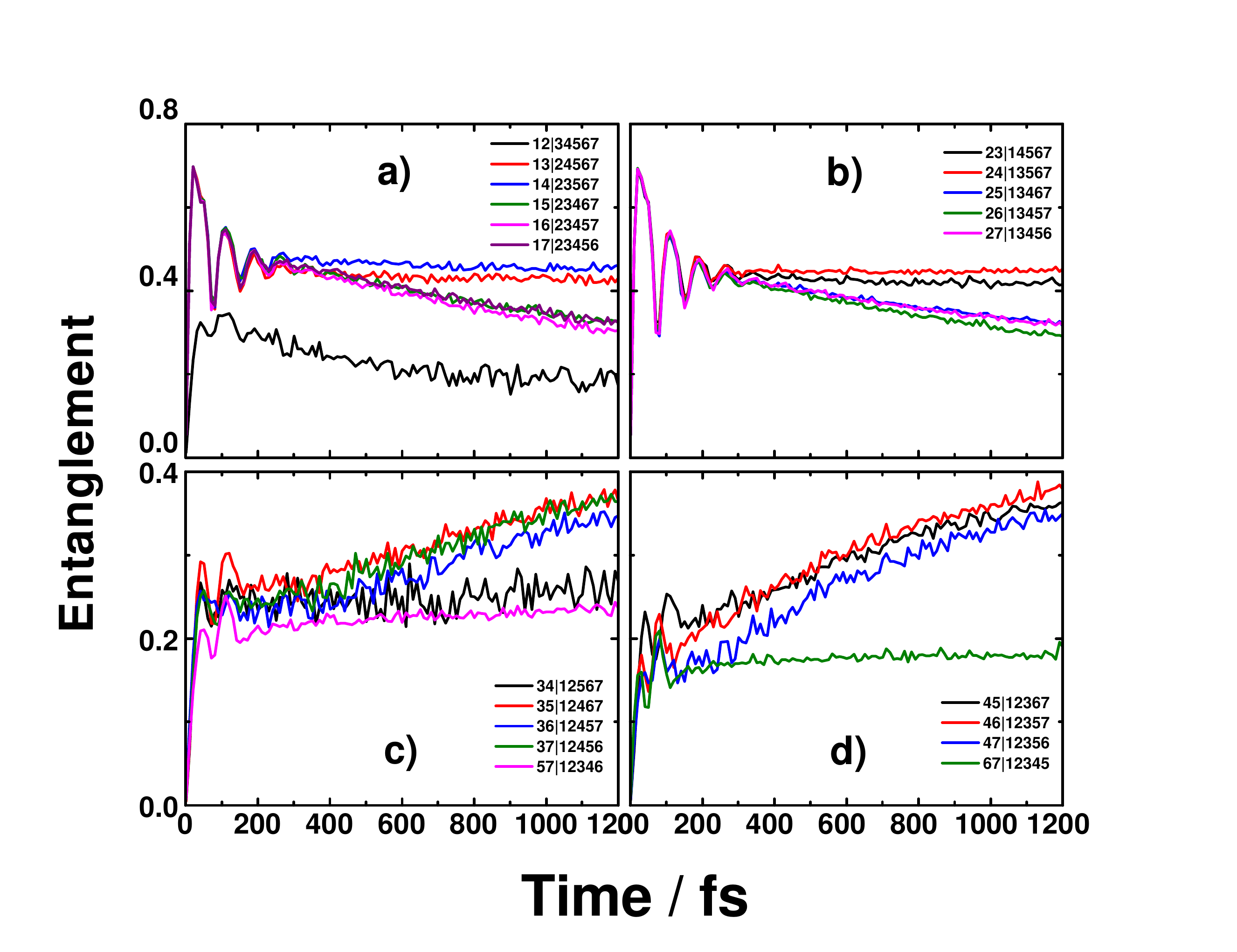}
\par\end{centering}
\caption{Entanglement measures $\sqrt{\eta_S}$ for the full FMO system at $77K$ with site one initially excited. These four plots show measures $\sqrt{\eta_S}$ computed across all $21$ bipartitions of the seven chromophore system into a pair of chromophores and the remaining quintuplet. Any measure that includes either chromophore $1$ or chromophore $2$ (but not both) on one side of the bipartition exhibits oscillations and the value of the measure is large. Any measure that has both chromophore $1$ and $2$ on the same side of the bipartition takes lower values and exhibits rapid growth in the first $100$ fs, but never exceeds $0.5$ in value. \label{doubles}}
\end{figure}

Figures~\ref{PairwiseS1},~\ref{bounds1} and~\ref{bounds2} give a picture of entanglement that is determined by the set of pairwise entanglements and the entanglement of single chromophores with the rest.  Even in this case we are seeing aspects of the mutipartite nature of entanglement in this system, as these measures refer to different partitions of the system. However, there are many more partitions that in general can exhibit multipartite entanglement structure. The measures $\sqrt{\eta_S}$ for all bipartitions of the seven chromophore system into a pair of chromophores and the other five are shown in Figure~\ref{doubles} for the FMO system at $77K$ with site one initially excited.  There are $21$ such partitions.

The measures $\sqrt{\eta_S}$ shown in Figure~\ref{doubles} exhibit two distinct types of behavior. Any measure $\sqrt{\eta_S}$ that is computed across a bipartition that separates sites $1$ and $2$ exhibits coherent oscillations and values of the measure that are large ($>0.5$). Any measure that is computed across a bipartition does not separate sites $1$ and $2$ has a rapid rise in the value of the measure initially by the value typically remains small $(<0.5)$.

\begin{figure}
\begin{centering}
\includegraphics[width=0.75\textwidth]{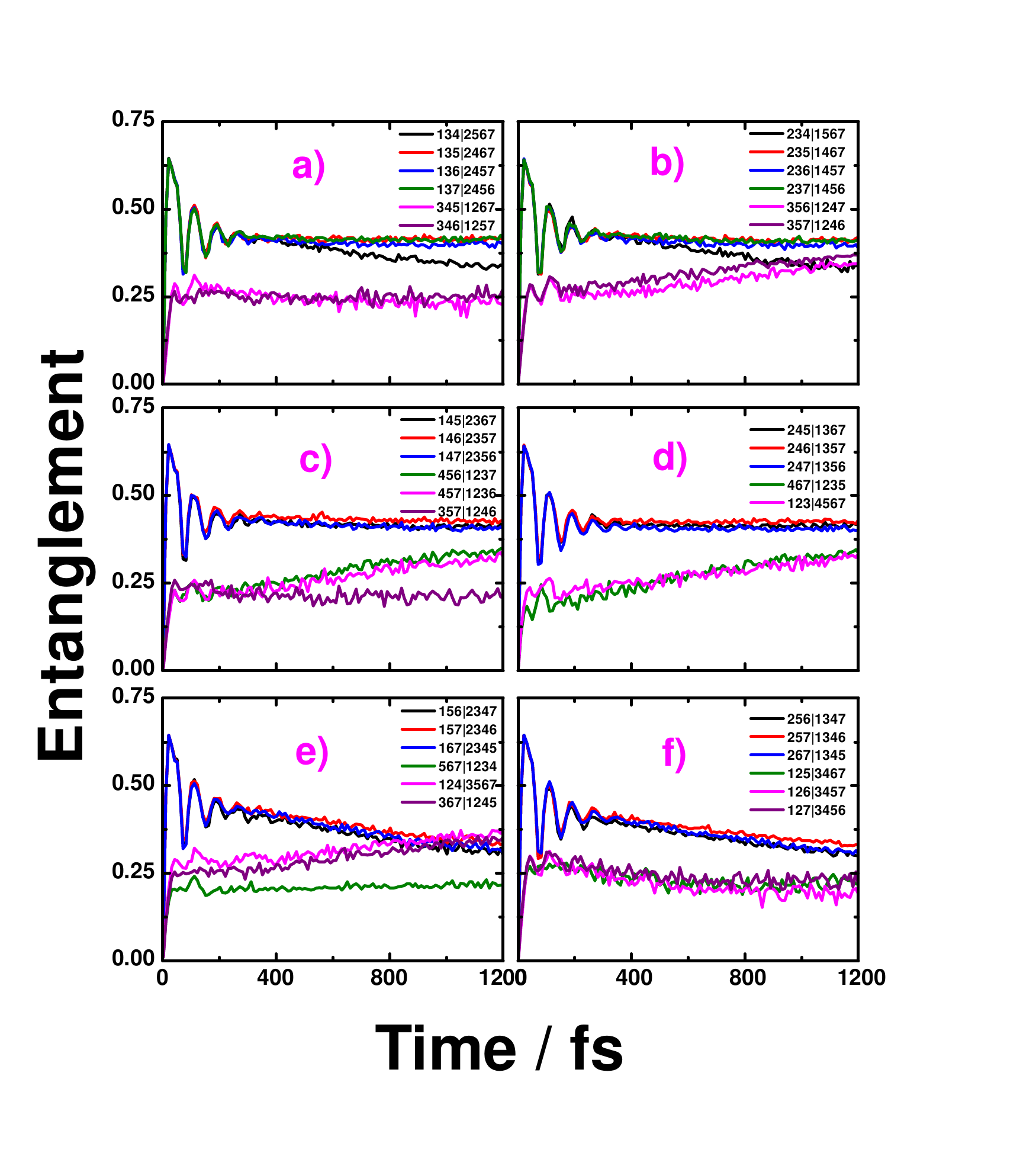}
\par\end{centering}
\caption{Entanglement measures $\sqrt{\eta_S}$ for the full FMO system at $77K$ with site one initially excited. These six plots show measures $\sqrt{\eta_S}$ computed across all $35$ bipartitions of the seven chromophore system into three of chromophores and the remaining four.  Any measure that includes either chromophore $1$ or chromophore $2$ (but not both) on one side of the bipartition exhibits oscillations and the value of the measure is large. Any measure that has both chromophore $1$ and $2$ on the same side of the bipartition takes lower values and exhibits rapid growth in the first $100$ fs, but never exceeds $0.5$ in value. \label{triples}}
\end{figure}

There remain further bipartitions of the seven chromophore FMO complex, namely those that divide the system into three chromophores and the remaining four. There are $35$ distinct bipartitions of this type, and the corresponding measures $\sqrt{\eta_S}$ are shown for the FMO complex at $77K$ in which site one is initially excited in Figure~\ref{triples}. The picture we obtain from Figure~\ref{triples} confirms that given by the previous measures displayed in Figures~\ref{bounds1},~\ref{bounds2} and~\ref{doubles}. Large ($>0.5$) values of the measures, and coherent oscillations, occur for any measure $\sqrt{\eta_S}$ computed across a bipartition that divided chromophore one from chromophore two. Any measure $\sqrt{\eta_S}$ computed across any bipartiton that does not separate chromophores one and two rises rapidly but remains small ($<0.5$) throughout the evolution.

\subsection{Beyond the single exciton manifold}

In addition to computing the measures of entanglement $\sqrt{\eta_S}$ above, which are based on simulations by the HEOM method in the one-exciton subspace, we wish to investigate what the effect of the presence of either zero excitons or more than one exciton in the system. We conducted a number of tests where we reinserted the ground state density matrix $\rho_0 = |0000000\rangle\langle0000000|$ and the two-exciton density matrix $\rho_2 = |0000011\rangle\langle0000011|$ in order to determine how the measures of entanglement would be affected. In the first test, we inserted the ground state $\rho_0$ on its own, yielding the following expression for the density matrix (where $\rho_1$ is the density matrix for the single-exciton subspace):
\begin{equation}
\rho = \frac{\rho_0+|\alpha|^2\rho_1}{1+|\alpha|^2}
\end{equation}

\begin{figure}
\begin{center}
\includegraphics[width=0.75\textwidth]{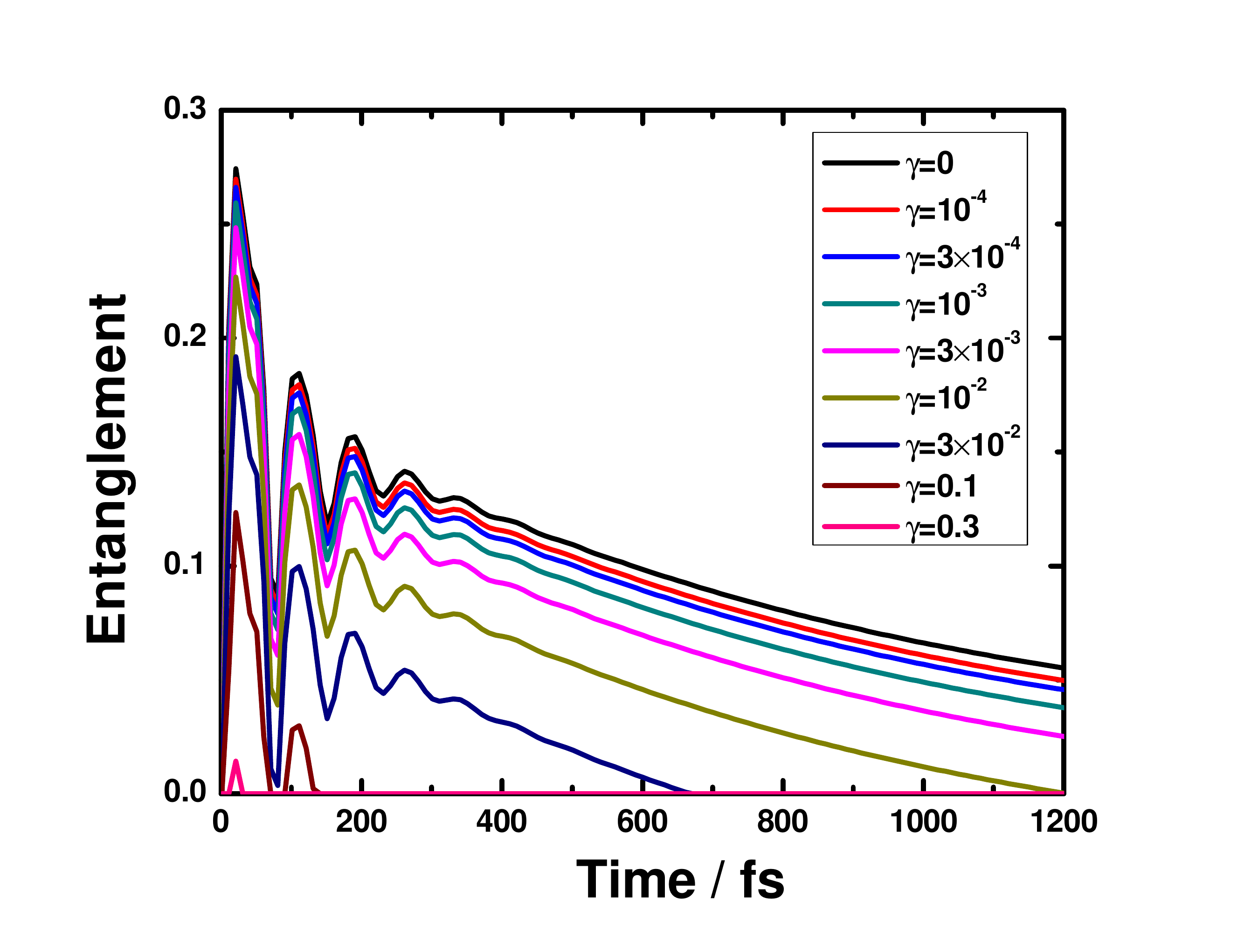}
\caption{ A comparison of the effects of adding in the two-exciton subspace $\gamma|\alpha|^2|0000011\rangle\langle0000011|$ for different values of $\gamma$, with $|\alpha|^2=0.5$. The concurrence between sites one and two is plotted for the density matrix in~\ref{zeroonetwo}, with $\rho_2=|0000011\rangle\langle0000011|$}
\label{TwoExciton1}
\end{center}
\end{figure}

\begin{figure}
\begin{center}
\includegraphics[width=0.75\textwidth]{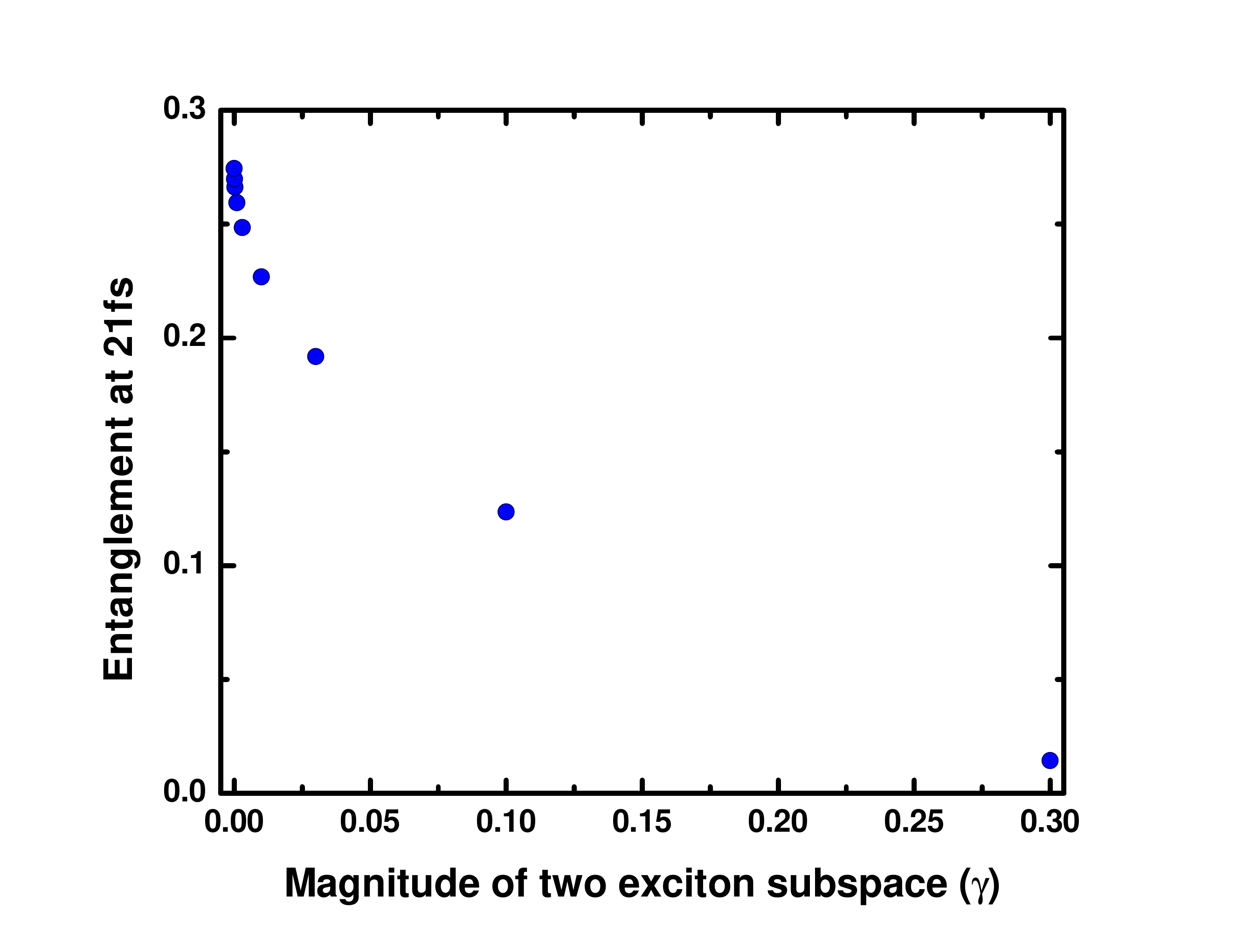}
\caption{A curve showing how the amplitude of the concurrence between sites 1 and 2 at 21 fs varies as a function of $\gamma$, with the density matrix from~\ref{zeroonetwo}, with $\rho_2 = |0000011\rangle\langle0000011|$ and $|\alpha|^2 = .5$.}
\label{TwoExciton2}
\end{center}
\end{figure}

In our second test, we added in the two-exciton subspace alone, without the ground state:
\begin{equation}\label{eqonetwo}
\rho = \frac{\rho_1+|\alpha|^2/2\rho_2}{1+|\alpha|^2/2}
\end{equation}
We then inserted $\rho_0$ and $\rho_2$ as follows:
\begin{equation}\label{zeroonetwo}
\rho = \frac{\rho_0+|\alpha|^2\rho_1+|\alpha|^4/2\rho_2}{1+|\alpha|^2+|\alpha|^4/2}
\end{equation}
When we added in both the vacuum state and the two-exciton subspace $|0000011\rangle\langle0000011|$ and varied $\alpha$, we found that for values as small as $|\alpha|^2 = .01$, the entanglement completely disappeared. We then experimented with adding in both the ground state and an exponentially decaying two-exciton subspace, $\rho_2 = e^{-\gamma t} |0000011\rangle\langle0000011|$, and, as expected, as $e^{-\gamma t}$ goes to zero, we recover some entanglement between sites 1 and 2, although the magnitude is still diminished by the presence of the vacuum state~\ref{gamma}. In order to get a sense of how quickly the entanglement recovers, we calculated the concurrence for the density matrix in equation~\ref{zeroonetwo}, which includes the ground state $|0000000\rangle\langle0000000|$ and the two-exciton subspace $|0000011\rangle\langle0000011|$ scaled by a factor $\gamma \in [0, 1]$:
\begin{equation}\label{gamma}
\rho = \frac{\rho_0+|\alpha|^2\rho_1+\gamma|\alpha|^4/2\rho_2}{1+|\alpha|^2+\gamma|\alpha|^4/2}
\end{equation}
The results are plotted in Figs. \ref{TwoExciton1} and \ref{TwoExciton2}.

\begin{figure}
\begin{center}
\includegraphics[width=0.75\textwidth]{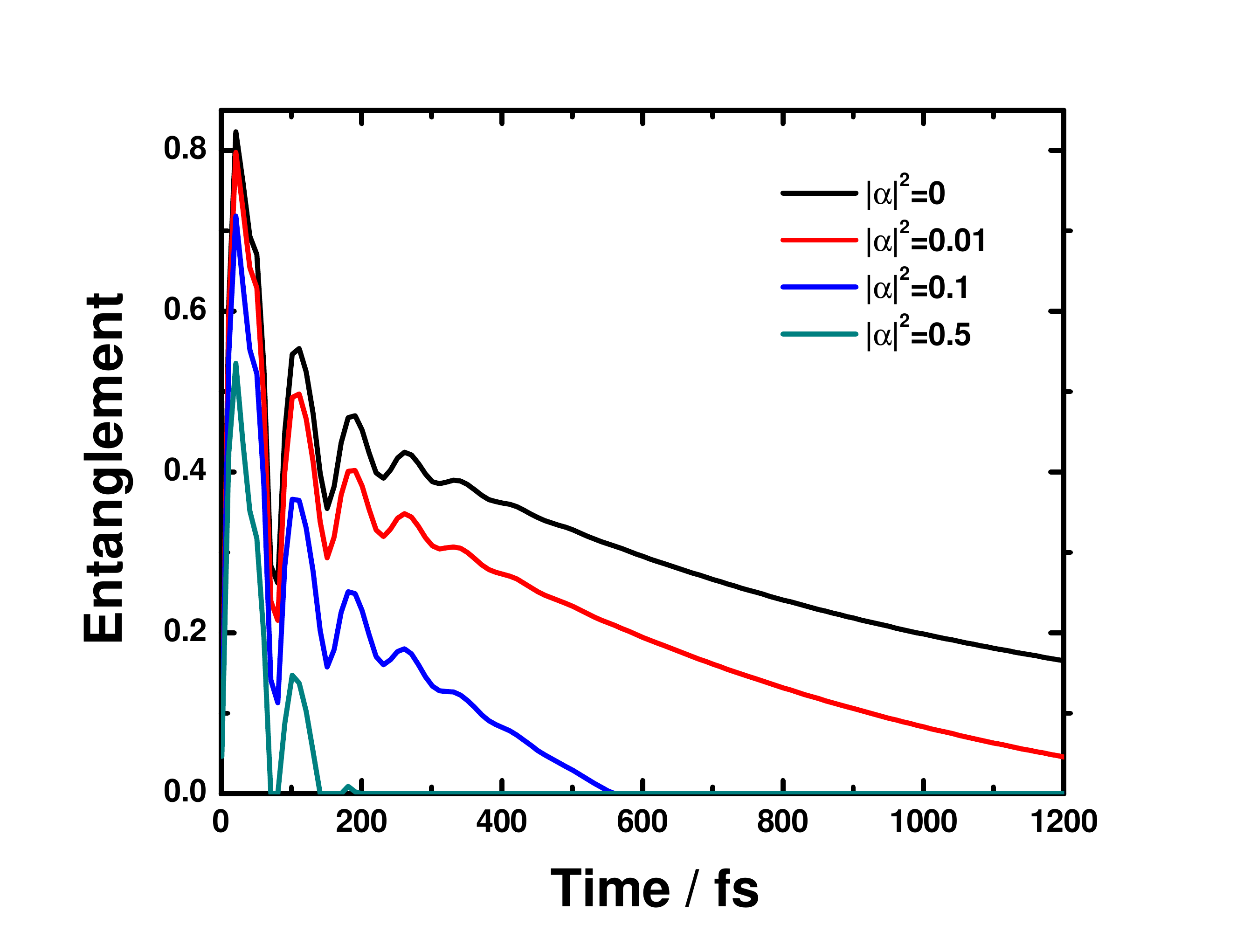}
\caption{A comparison of the effects of adding in the two-exciton subspace for different values of $|\alpha^2|$. The concurrence between sites one and two is plotted for the density matrix in equation~\ref{eqonetwo}, with $|\alpha^2|= .5, .1, .01$.}
\label{onetwo}
\end{center}
\end{figure}

\section{Conclusions}

In summary, we used the direct computation of the convex roof to calculate the evolution of  number of bipartite entanglement in the FMO complex
via the scaled HEOM approach.  For the simulations in which site $1$ is initially excited, the dominant pair is site $1$ and $2$, while in the cases where $6$ is initially excited site $5$ and $6$ are most entangled. This indicates that entanglement is dominant in the early stages of exciton transport, when the exciton is initially delocalized away from the injection site. In addition we observe that the entanglement mainly happens among the sites involved in the pathway. For the site $1$ initially excited case, the entanglement of site $5$, $6$ and $7$ is almost zero. For the site $6$ initially excited situation, there is seldom entanglement for site $1$ and $2$.

Although the final state is the same for both initial conditions, the role of site $3$ and site $4$ during the time evolution is different. For the initial condition where site $1$ is excited, the entanglement is transferred to site $3$ and then from site $3$ to site $4$. While for the site $6$ initially excited case, sites $4$ and $5$ first become entangled with site $6$ and then sites $3$ and $4$ become entangled. This is due to the fact that site $3$ has strong coupling with site $1$ and $2$, while site $4$ is coupled more strongly to sites $5$, $6$ and $7$.

The initial condition plays an important role in the entanglement evolution, the entanglement decays faster for the cases where site $6$ is initially
excited compared with cases where the site $1$ is initially excited. This is consistent with recent models that include the nature of the excitation caused by the incident light, and which show a strong dependence of the amount of entanglement generated on the details of the excitation process~\cite{addthree}.
Increasing the temperature unsurprisingly reduces the amplitude of the entanglement
and also decreases the time for the system goes to thermal equilibrium, in agreement with prior work.

Most entanglement measures computed previously for FMO were chosen on the basis of ease of calculation. The negativity and logarithmic negativity are straightforward to compute for all states~\cite{Thorwart:2009p8612,Caruso:2010p8915}. The global and bipartite relative entropy of entanglement can be made straightforward to compute by restriction to the single exciton subspace~\cite{Bradler:2010p7069,Sarovar:2010p6945}. The bipartite concurrence and tangles can be computed easily for pairs of chromophores~\cite{Sarovar:2010p6945,Fassioli:2010p8617}. In all cases the chosen measures of simplifications thereof enable one to avoid computing the convex roof over different ensembles representing a mixed state. In this paper we explored the difficulty of such calculations, and find that measures that yield the bipartite entanglement across cuts of 3,4, and 5 qubit subsystems may be computed with modest effort. We computed monogamy bounds to obtain a lower bound on a number of measures and the convex roof to obtain an upper bound. The closeness of these two bounds gives a measure of how well the convex roof is performing. For pure states in the single exciton manifold the monogamy bounds are saturated~\cite{SanKim:2008p8940} - however this is not knwon to be the case for the mixed states of interest here. The convex roof technique enables us to extend the set of measures that have been computed for FMO, and also shows that the computation of entanglement for this system is not restricted by the difficulty of the convex roof procedure. This procedure could also be used, with no increase in computational cost, to analyze entanglement in multiexcitonic models.

For the full system of seven chromophores it was necessary to restrict the convex roof optimization to the single exciton subspace in order to make the calculations tractable. We performed a complete calculation of measures $\sqrt{\eta_S}$ across all $63$ bipartitions, which contains all information concerning the multipartite entanglement present in the system. The results of these calculations for site one initially excited confirm the conclusions of calculations on smaller subsystems: the structure of entanglement in this system can be understood in terms of pairwise entanglement. The fact that the other measures of entanglement add no new information to the picture is perhaps suprising. It remains to be seen whether this is a general (but currrently unproven) property of the single excitation subspace, or whether it is a property of the particular dynamics of the FMO system. We leave these questions to future investigations.

\section{Acknowledgment}
This Project is supported by NSF CCI center, "Quantum Information for Quantum Chemistry(QIQC)",
 Award number CHE-1037992, and by NSF award PHY-0955518.

\bibliographystyle{aipnum4-1}
\bibliography{Entanglement}

\appendix

\subsection{The Cayley Map}
The Cayley map is a self-inverse map from the algebra $u(N)$ to the group $U(N)$. The Cayley map is a map between a number of Lie algebras and their respective groups. It was introduced as a map from $so(N)$ to $SO(N)$~\cite{Cayley1846}. The Cayley map is defined by
\begin{equation}\label{Intro1}
\text{Cay}(a)= A = \paren{I-a}\paren{I+a}^{-1}
\end{equation}
where $a$ is an element of the algebra being considered, and $A$ is an element of the group. Likewise, we have
\eq{Intro2}{\text{Cay}{A}=a=\paren{I-A} \paren{I+A}^{-1}}

In the case of the unitary group, the Cayley map is a bijection between $u(N)$ and the set $U(N)-\mathscr{E}$, where $\mathscr{E}$ is the set of ``exceptional elements.'' $\mathscr{E}$ is the set of all elements $A$ such that $I+A$ is singular, and can be characterized as the set of all elements $A$ with at least one eigenvalue $-1$. The exceptional elements on $SO(3)$ are the reflections. For all such elements $E$, $I+E$ has a 0 eigenvalue, and is not invertible, so the Cayley map is not defined on these elements; however, this will not hinder our attempts to minimize $\eta$ over $U(N)$. Since we are performing numerical optimization, we only care that we can get arbitrarily close to a given local optimum. The closure of the image of the Cayley map on $u(N)$ is all of $U(N)$, so we will still be able to identify minima located at exceptional points.

Because $u(N)$ is easily parameterized by $N^2$ parameters, we can therefore parameterize $U(N)$ by $N^2$ parameters via the Cayley map. Given a set of $N^2$ parameters $\{p_1,\ldots,p_{N^2}\}$, the corresponding element of $U(N)$ is then:
\eq{Intro3}{A = \text{Cay}(a(p_1,\ldots,p_{N^2}))}
where $a$ is the element of $u(N)$ given by the parameters $p_i$ under a standard parametrization. In the current work we use the basis of tensor products of Pauli matrices for the algebra $su(N)$. The virtue of the Cayley map is that it gives us an easily understood and easily implemented way to parameterize $U(N)$. The Cayley map thus provides somewhat simpler parametrization than that used in prior work on the convex roof optimization in~\cite{Rothlisberger:2009p7007}. Comparison of the performance of our method with the simulated annealing approach described in Appendix B of~\cite{Zyczkowski:1999p2567} shows a substantial advantage to parametrization by the Cayley map combined with steepest descent. We leave detailed comparison of our method with that of~\cite{Rothlisberger:2009p7007}, and the evaluation of other optimization techniques beyond steepest descent, to future work.


\end{document}